\documentclass[oldversion]{aa} 
\usepackage{graphicx}
\usepackage{longtable}
\usepackage{txfonts}
\usepackage{rotating}
\usepackage{lscape}
\newcommand{\gsim}{\;\lower.6ex\hbox{$\sim$}\kern-7.75pt\raise.65ex\hbox{$>$}\;}
\newcommand{\lsim}{\;\lower.6ex\hbox{$\sim$}\kern-7.75pt\raise.65ex\hbox{$<$}\;}

\begin{document}
\title{Terzan 8: a Sagittarius-flavoured globular cluster\thanks{Based on 
observations collected at ESO telescopes under 
programme 087.B-0086}\fnmsep\thanks{
   Table 2 is only available in electronic form at the CDS via anonymous
   ftp to {\tt cdsarc.u-strasbg.fr} (130.79.128.5) or via
   {\tt http://cdsweb.u-strasbg.fr/cgi-bin/qcat?J/A+A/???/???}}
 }

\author{
E. Carretta\inst{1},
A. Bragaglia\inst{1},
R.G. Gratton\inst{2},
V. D'Orazi\inst{3,4},
S. Lucatello\inst{2},
\and
A. Sollima\inst{1}
}

\authorrunning{E. Carretta et al.}
\titlerunning{Abundance analysis in Terzan 8}

\offprints{E. Carretta, eugenio.carretta@oabo.inaf.it}

\institute{
INAF-Osservatorio Astronomico di Bologna, Via Ranzani 1, I-40127 Bologna, Italy
\and
INAF-Osservatorio Astronomico di Padova, Vicolo dell'Osservatorio 5, I-35122
 Padova, Italy
\and
Dept. of Physics and Astronomy, Macquarie University, Sydney, NSW, 2109 Australia 
\and
Monash Centre for Astrophysics, Monash University, School of Mathematical 
Sciences, Building 28, Clayton VIC 3800, Melbourne, Australia
  }

\date{}

\abstract{Massive globular clusters (GCs) contain at least two generations of 
stars with slightly different ages and clearly distinct light elements 
abundances. The Na-O anticorrelation is the best studied chemical 
signature of multiple stellar generations. Low-mass clusters
appear instead to be usually chemically homogeneous. We are investigating 
low-mass GCs to understand what is the lower mass limit where multiple 
populations can form, mainly using the Na and O abundance distribution.
We used VLT/FLAMES spectra of giants in the low-mass, metal-poor GC Terzan~8,
belonging to the Sagittarius dwarf galaxy, to determine abundances of  Fe, O, Na,
$\alpha-$, Fe-peak, and neutron-capture elements in six stars observed with UVES
and 14 observed with GIRAFFE. The average metallicity  is [Fe/H]$=-2.27\pm0.03$
(rms=0.08), based on the six high-resolution  UVES spectra. Only one star,
observed with GIRAFFE, shows an enhanced  abundance of Na and we tentatively
assign it to the second generation. In this  cluster, at variance with what
happens in more massive GCs, the second generation seems to represent at most a
small minority fraction. We discuss the  implications of our findings, comparing
Terzan~8 with the other Sgr dSph  GCs, to GCs and field stars in the Large
Magellanic Cloud,  Fornax, and  in other dwarfs galaxies.
}

\maketitle

\section{Introduction}

Once considered good examples of simple stellar populations, Galactic globular
clusters (GCs) are currently recognised as having formed in a complex chain of 
events occurring very early in their lifetimes. The fossil record of these
events is encrypted in the chemical composition of different stellar populations
left over by the process of cluster formation.
Exploiting the largest and most homogeneous dataset available up to date, 
Carretta et al. (2010a) showed that most, perhaps all GCs host multiple stellar
populations that can be traced by variations of Na and O abundances. First
extensively studied by the Lick-Texas group (see Kraft 1994), these variations are found to
be anti-correlated with each other (see, e.g., Gratton, Sneden \& Carretta 2004,
and Gratton, Carretta \& Bragaglia 2012 for recent reviews). 
This notion is corroborated and extended by photometry, showing that many GCs
present spreads or even splits of the evolutionary sequences (see, e.g., Lee et
al. 2009, Carretta et al. 2011a, Milone et al. 2012a, 2013, Piotto et al. 2012,
and references therein). These are attributed to the effect of different
chemical composition, in particular of light elements like He, C, N, and O
(e.g., Carretta et al. 2011b, Sbordone et al. 2011, Milone et al. 2012a). 

Our ongoing FLAMES survey (see Carretta et al. 2006, 2009a, 2009b) allowed for
the first time a quantitative estimate of a few relevant parameters.
We noted that all GCs analysed have stars of composition both primordial
(the first generation, FG) and modified (the second generation, SG), and that
the latter are always the majority (with a $\sim30-70$ \% proportion between FG
and SG, Carretta et al. 2009a). We further noted that the extension of the Na-O
anticorrelation is well correlated to cluster mass (Carretta et al. 2010a). 
Moreover, there seems to be a sort of {\em observed} minimum cluster mass for 
the appearence of a Na-O anticorrelation, i.e., of a second generation of 
stars within a cluster. This could be due to a real effect (see Bekki 2011 and 
also Caloi \& D'Antona 2011) or to the scarce statistics available for low-mass
clusters. After our survey of more than 20 populous GCs (see, e.g., Carretta et
al. 2009a,b) we and others are trying to sistematically sample also the
border between the lower mass end of the GC population and the higher mass end
of old open clusters (see e.g. Bragaglia et al. 2012 and Geisler et al.
2012, and Sect.~5.1). Our intent is to study the chemical composition of a
large, representative sample of objects in this mass region, to give robust
contraints to models of cluster formation and evolution. 

It is also important to study clusters belonging to other galaxies. A Na-O
anticorrelation has been found in old Large Magellanic Cloud (Mucciarelli et al.
2009) and Fornax (Letarte et al. 2006) clusters. However, there are closer GCs
of extragalactic origin, namely those of the disrupting Sagittarius dwarf galaxy
(Ibata, Gilmore \& Irwin 1994), for which the situation is less clear. 
According to Ibata, Gilmore \& Irwin (1995) and Da Costa \& Armandroff (1995),
at least four GCs appear to be associated with the Sgr dSph, namely M54, 
Arp~2,  Ter~7, and Ter~8. For Law \& Majewski (2010), genuine Sgr GCs
are Arp~2, M~54, NGC~5634, Ter~8, and Whiting~1, with  NGC~5053, Pal~12, and
Ter~7 (plus the open cluster Be~29) being likely members; other authors may differ
on the list, but this is irrelevant for the present paper. Apart from the
massive cluster M~54, for which we obtained FLAMES spectra of about 100 stars of
the cluster and surrounding Sgr field (Carretta et al. 2010b,c), the other Sgr
GCs for which chemical abundances based on high-resolution spectroscopy are
available are all low-mass GCs. A few stars were analysed in each of them,
namely four in Pal~12 (Cohen 2004), with $M_V=-4.48$ (from the Harris 1996
catalogue), five in Ter~7 (Tautvaisiene et al. 2004, Sbordone et al. 2007;
$M_V=-5.05$), two in Arp~2 ($M_V=-5.29$), and three in  Ter~8 (Mottini,
Wallerstein \& McWilliam 2008; $M_V=-5.05$). No significant spread in
proton-capture elements were found.  The samples are admittedly small, but
these clusters are distant (about 20-30 kpc from the Sun) and even the (often
few) stars near the red giant branch (RGB) tip are rather faint for
high-resolution spectroscopy. A possible exception is NGC~4590 (closer to the
Sun and for which Fe, Na, and O were presented for more than 40 stars in
Carretta et al. 2009a,b), but this cluster is not unanimously assigned to the
Sgr family of GCs.

Ter~8 has been studied by Mottini et al. (2008), together with the other Sgr GC
of very low metallicity, Arp~2. They obtained high-resolution spectra  using the
MIKE spectrograph  at the 6.5m Magellan telescope, measuring metallicity,
$\alpha$-elements, and heavy elements. Unfortunately, they measured O, but not
Na. The abundance of O looks rather homogeneous in both clusters: 
[O/Fe]=0.22\footnote{We adopt the usual spectroscopic notation, $i.e.$  [X]=
log(X)$_{\rm star} -$ log(X)$_\odot$ for any abundance quantity X, and  log
$\epsilon$(X) = log (N$_{\rm X}$/N$_{\rm H}$) + 12.0 for absolute number density
abundances.} (rms 0.04) dex for Arp~2 and 0.64 (rms 0.18) dex for
Ter~8.\footnote{These values are obtained using Fe {\sc ii} to convert [O/H] to
[O/Fe]. Looking at Tables 5 and 6 of Mottini et al. (2008), they used Fe {\sc
ii} for Arp~2 and Fe {\sc i} for Ter~8. No relevant difference comes from this
choice.} 

\begin{figure}
\centering
\includegraphics[scale=0.45]{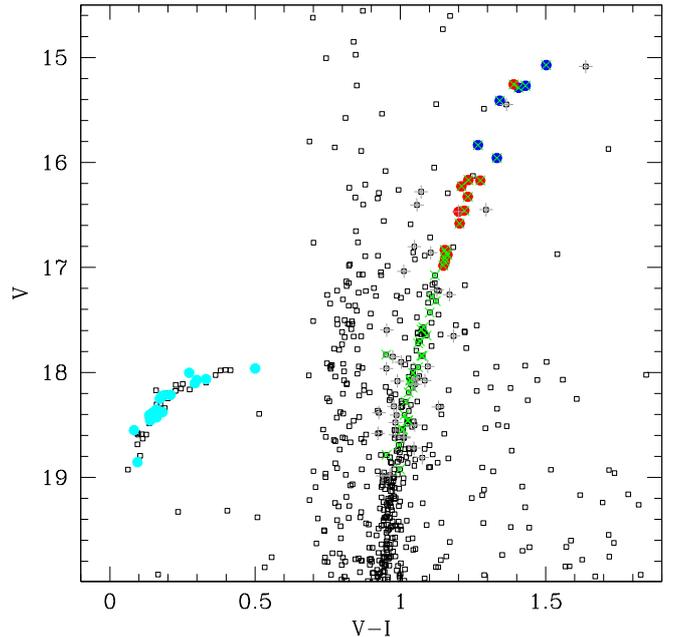}
\caption{$V,V-I$ colour-magnitude diagram (CMD) of Terzan 8 from Montegriffo et
al. (1998; open symbols). Filled circles indicate the stars in our sample
observed with FLAMES/UVES (in blue) and with FLAMES/GIRAFFE (in red) and used
for the abundance analysis (with $V<17$). Filled cyan symbols are HB stars, for
which we could not measure the RV. The HB star 1350, at $V-I=0.5$, is a known RR
Lyrae (V2, of type RRc, according to Salinas et al. 2005). Green crosses
indicate members on the basis of their RV; grey plus signs are for non members.}
\label{f:cmdm08vi}
\end{figure}

Our paper adds six more stars for which the complete set of elements could be
obtained, and 10 for which at least Fe and Na could be derived; these are nearly
all the stars on the RGB brighter than $V=17$ (see Fig.~\ref{f:cmdm08vi}). The
paper is organized as follows: our observations are presented in Sect. 2 and the
abundance analysis in Sect. 3, while the results are illustrated in Sect. 4.
Sects.5 and 6 are devoted to discuss and to summarise our findings,
respectively.

\section{Observations}\label{obs}

\subsection{Photometry and cluster parameters}\label{obsphot}

We used the $V,I$ photometry by Montegriffo et al. (1998), kindly  provided by
the authors, to select our targets. 
The data were obtained with EMMI@NTT, on a field of view of
$9.15\times8.6$ arcmin$^2$ and calibrated using a shallower set of data obtained
at the 0.9m CTIO telescope (see Montegriffo et al. 1998 for details). Optical
magnitudes were complemented, whenever possible,  with $K$ band magnitudes  from
the 2MASS Point Source Catalogue (Skrutskie et al. 2006).

The $V,V-I$ colour magnitude diagram (CMD) of Ter~8 is shown in
Fig.~\ref{f:cmdm08vi}, with different symbols for the sample of stars observed
with FLAMES and according to their radial velocities (RV) measured on our
spectra (see below). The tidal radius $r_t$ is less than 4\arcmin, from
the Harris (1996) catalogue, while the half light radius $r_h$ is 0.85\arcmin;
these values where revised by Salinas et al. (2012) to $r_t$=5.56\arcmin \ and
$r_h$=2\arcmin. With both sets of values, Ter~8 is a small cluster and it is not
easy to select targets for a multi-object spectrograph like FLAMES, which has a
field-of-view diameter of 25\arcmin. Furthermore, Ter~8 is quite far, at about
26 kpc from the Sun, so only a small fraction of the cluster stars can provide
high S/N spectra in reasonable observing times. We selected the targets from the
photometry files, after astrometrisation to the GSC-2 systems using programs
written by P. Montegriffo at the Bologna Observatory.  As in our past works of
this project, all the targets are  free from neighbours; they have no star
closer than 3\arcsec \ (or 2\arcsec \ but  only if at least 2 mag fainter) and
were selected to lie near  the RGB (or Horizontal Branch, HB) ridge line.

\subsection{FLAMES data}\label{obsspec}

We obtained exposures with the multi-object spectrograph FLAMES@VLT 
(Pasquini et al. 2002), as in our  previous works on the Na-O 
anti-correlation (see e.g. Carretta et al. 2009a,b; Bragaglia et al. 2012). 
The observations were obtained in service mode (see Table~\ref{t:logobs} for a
log). We used the UVES 580nm setup  ($\lambda\lambda\simeq4800-6800$~\AA) and
the GIRAFFE high-resolution gratings HR11 (containing the Na~{\sc i} doublet at
5682-5688~\AA) and HR13 (which contains the forbidden [O {\sc i}] line at 
6300~\AA\  and the Na~{\sc i}  doublet at 6154-6160 \AA). 

\begin{table}
\centering
\caption{Log of the observations}
\setlength{\tabcolsep}{1mm}
\begin{tabular}{ccccccc}
\hline\hline
OB  & UT date & UT$_{init}$ & exptime & airmass &seeing &HR\\
       & (Y-M-D) & (h:m:s)      & (s) &         &(arcsec)&\\
\hline
A &2011-06-24 &06:40:15.900 &4720 &1.018 &1.90 &11 \\
B &2011-06-24 &08:04:10.749 &4720 &1.106 &1.78 &11 \\
C &2011-06-28 &03:46:57.561 &4720 &1.159 &1.20 &11 \\
D &2011-06-28 &05:15:05.685 &4720 &1.029 &1.45 &11 \\
E &2011-07-04 &02:27:08.284 &3455 &1.346 &2.21 &11 \\
F &2011-08-27 &01:08:41.704 &5400 &1.037 &1.11 &11 \\
G &2011-07-22 &05:57:38.439 &4720 &1.079 &1.66 &13 \\
H &2011-08-27 &02:56:06.129 &4720 &1.034 &1.30 &13 \\
I &2011-08-28 &01:57:43.600 &4720 &1.013 &1.96 &13 \\
L &2011-08-28 &03:22:55.549 &4720 &1.066 &1.23 &13 \\
M &2011-08-30 &02:37:19.319 &4720 &1.028 &1.38 &13 \\
N &2011-08-31 &03:44:36.510 &4720 &1.121 &0.70 &13 \\
\hline  		       
\end{tabular}		 
\label{t:logobs} 	 
\end{table}		 

As shown in Fig.~\ref{f:cmdm08vi}, we selected seven among the brightest RGB
stars for the UVES fibres ($R\simeq45000$). The GIRAFFE fibres (at
$R\simeq24200$ for HR11 and 22500 for HR13) were  allocated to fainter objects
on/near the RGB or the HB. These stars, mostly fainter than $V\simeq17$, were 
observed to determine their RV (i.e., their membership status). However, the
spectral regions in HR11 and 13 are not ideal for the HB stars, especially at
this low metallicity and S/N, and we could not measure their RV.  Information on
all observed  stars can be found in Table~\ref{t:tabphot} (only available in
electronic form).  We give: ID; equatorial coordinates; $V$, $I$, and 2MASS
(Skrutskie et al. 2006) $K$ magnitudes; and the  heliocentric RV with its error.

The spectra were reduced (bias and flat field corrected, 1-D extracted, and
wavelength  calibrated)  by the ESO personnel. We applied sky subtraction and
division by an  observed early type star (UVES) or a synthetic spectrum
(GIRAFFE) to correct for  telluric features near the [O {\sc i}] line, using the
IRAF\footnote{IRAF is  distributed by the National Optical Astronomical
Observatory, which are operated by  the Association of Universities for Research
in Astronomy, under contract with the  National Science Foundation.} routine
{\em telluric}. The latter correction was  applied only to the UVES and bright
GIRAFFE samples. 
We shifted all spectra according to the heliocentric velocity and combined the
individual exposures; the stars heliocentric RVs were then measured  using 
{\em rvidlines} in IRAF.  The UVES final spectra have S/N in the range 45-80,
the GIRAFFE spectra of stars retained for abundance analysis have median S/N
values of 60 and 33 for the HR11 and HR13 spectra, respectively.

\begin{figure}
\centering
\includegraphics[bb=50 170 365 710, clip,scale=0.75]{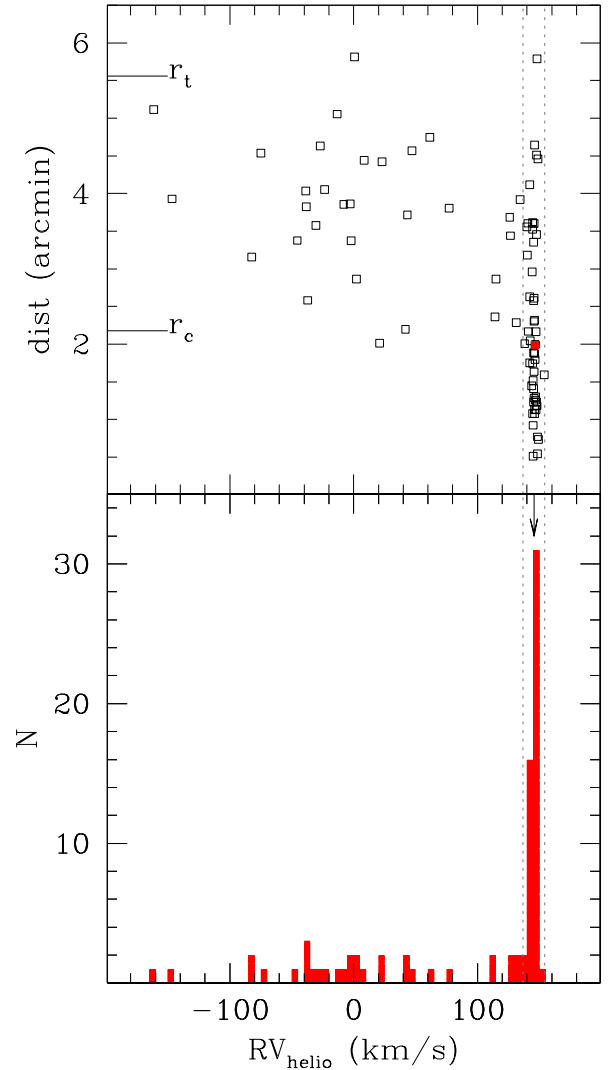}
\caption{Upper panel: RVs versus distance from the cluster centre. The core and
tidal radii, from Salinas et al. (2012) are indicated. The red, filled point is
the high-Na star 2023 (see text). Lower panel: histogram of all RVs for the
FLAMES spectra. The arrow indicates the average cluster RV, at about 145.5
km~s$^{-1}$. The two vertical lines in both panels indicate $\pm 3\sigma$ from
the average.} 
\label{f:rvter8} 
\end{figure}

The cluster average RV was computed separately for UVES and GIRAFFE spectra, to
check possible offsets due to the different resolution and wavelegth
calibration. However, we found the same average RV for both samples: 
$\langle RV_{UVES} \rangle=145.18$ (rms=0.69) km~s$^{-1}$ and 
$\langle RV_{GIRAFFE} \rangle =145.48$  (rms=2.90) km~s$^{-1}$.
Fig.~\ref{f:rvter8} shows the RV  distribution of our stars as a function of
distance from the cluster centre and the relative histogram; the cluster's
signature is evident in both panels. We considered as stars of Ter~8 the objects
with RV within 3$\sigma$ from the mean of the GC.
As expected, most of the cluster members are  centrally concentrated, the field
outliers are found outside about $1~r_c$, where the contamination is about 50\%.

As a comparison, Harris (1996) has $RV=130\pm8$ km~s$^{-1}$, taken originally 
from Da Costa \& Armandroff (1995), who observed four stars in the IR Ca triplet
region at a resolution of about 3~\AA, i.e., significantly lower than ours,
obtaining individual RVs of 123, 145, 121, and 185  km~s$^{-1}$. Formally, the
average of all four is in perfect agreement with our value; the lower value
cited in the catalogue is for the three most probable members. Given the very
small sample and the lower resolution, their value is in reasonable agreement
with ours. We cannot compare our values to Mottini et al. (2008) because they do
not provide the RVs for the stars they observed.

After pruning the sample using RVs we found a total of 53 member stars -in
particular six out of the seven with UVES spectra- and 29 non members out of a total of 101
targets observed  (for 19 spectra we could not measure a RV because of the very
low S/N and/or the not favourable wavelength range, as for the HB stars). The
individual RV values are given in Table~\ref{t:tabphot}; their error is typically
0.5 and 1.5 km~s$^{-1}$ for UVES and GIRAFFE, respectively. Ours is the first
measurement of the velocity dispersion in Ter~8; a discussion of the implication
of such a small dispersion (2.9 km~s$^{-1}$, for the larger GIRAFFE sample) 
for the cluster characteristics is deferred to a
dedicated paper (Sollima et al., in preparation). Here we wish to remark the
narrowness of the RGB sequence once the non-members are eliminated (see
Fig.~\ref{f:cmdm08vi}). 

No star noticeably far from the RGB ridge line survived the RV test, except for
star 36, which lies slightly to the blue of the RGB. It is at the level expected
for the red HB (RHB); however,  all the HB is blue in this cluster. Star 36 was
not analysed, but its spectrum does not show any peculiarity (such as, e.g.,
evident rotation, or evidence of binarity from the RVs). Compared to other stars
of  similar magnitude, this object looks slightly hotter. It could be a binary
with a bluer component (a white dwarf). Emanuele Dalessandro kindly checked  the
GALEX data (Schiavon et al. 2012) and the star is visible in the near UV, but
not on the far UV image, so there is no definitive evidence in favour of a
hotter component.  This star is a member on the basis of its RV, but is located
far from the centre, near the cluster $r_t$, so it could also be a field
interloper. 
Another interesting possibility is that we have found a post Blue Straggler star
(post-BSS), given its position in the CMD (see Renzini \& Fusi Pecci 1988, Fusi
Pecci et al.  1992). Some tentative identifications of post-BSS exist (see e.g.,
Ferraro \& Lanzoni 2013 for a  review), and we have found a few other candidates
in our studies of the HB (see e.g.  Gratton et al. 2013). However, we can not
confirm on the basis of the present data the true nature of star 36. The Ba 
{\sc ii} 6141~\AA\ line looks slightly more evident than for other stars of
Ter~8 of similar atmospheric parameters, and maybe slightly larger, as expected
for a  post-BSS, but the comparison is difficult, given the low S/N of the
spectra.

\section{Atmospheric parameters, abundance analysis, and metallicity}

We derived the abundances of several elements only for the 16 brightest member
stars.
The analysis followed as closely as possible the technique used in previous
studies by our group concerning the Na-O anticorrelation in GCs (see e.g. 
Carretta et al. 2006, 2009a,b for extensive references and methods). The main
difference was the use also of Na D lines to avoid having upper limits in Na,
due to the combination of low metallicity and S/N. In turn, we were forced to
abandon the usual estimates of the microturbulent velocity $v_t$ made by
minimizing the slope of the Fe~{\sc i} abundances as a function of the expected
line strengths (Magain 1984). The values of $v_t$ obtained with this technique
are unsuitable when applied to strong lines such as the Na D or the Ba lines,
resulting in strong trends as a function of the microturbulent velocity. The
adopted changes in the present abundance analysis are described below.

Line lists, atomic parameters, and reference solar abundances are taken
from Gratton et al. (2003). We measured EWs with the software Rosa (Gratton
1988), as described in Bragaglia et al. (2001).
We adopted the same automatic procedure for the definition
of  the local continuum around each line of previous papers, but we did not
correct the  GIRAFFE EWs to the UVES system because we did not have any star 
observed with both GIRAFFE and UVES.

\begin{figure}
\centering
\includegraphics[bb=102 146 428 705, clip,scale=0.65]{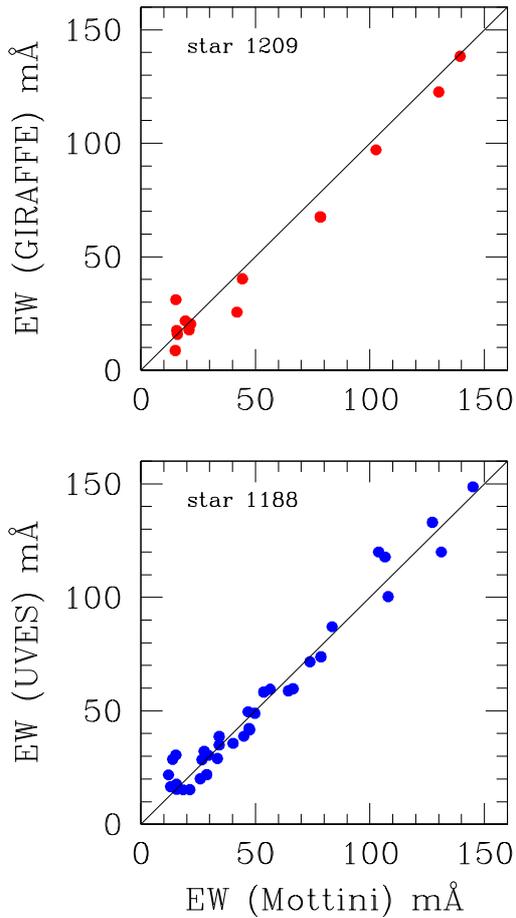}
\caption{Top panel: comparison of our $EW$s measured on the GIRAFFE spectrum of
star 1209 with those measured by Mottini et al. (2008). Lower panel: the same
for star 1188; in this case our $EW$s are measured on the UVES spectrum of the
star. In both panels, the solid line is the equality line.}
\label{f:ewum8}
\end{figure}

However, we have two stars in common with those observed by Mottini et al.
(2008) in Ter~8 by using the Magellan MIKE spectrograph, with a resolving
power of 40,000. Star 1188 (star 305 in Mottini et al.) and star 1209 (star 325
in Mottini et al.) were observed with UVES and GIRAFFE, respectively, in our
study. In Fig.~\ref{f:ewum8} we compare our measured $EW$s with those published
by Mottini et al. (2008). The agreement is good in both cases, the average
differences (in the sense us minus Mottini et al.) being  
$+0.7\pm 1.2, \sigma=6.9~m\AA$ (35 lines) and 
$-3.3\pm 2.0, \sigma=7.7~m\AA$ (14 lines) for
stars 1188 (=305) and 1209 (=325),  respectively.

Following the homogeneous approach used in previous papers (see e.g. Carretta et
al. 2009a,b) initial temperatures were derived from $V-K$ colours. We adopted
the calibration by Alonso et al. (1999, 2001) for effective temperatures and
bolometric corrections. Final temperatures were obtained through a relation with
$K$ magnitudes to minimize the star-to-star internal errors. Surface gravities
were derived from these temperatures and the position of stars on the CMD,
adopting reddening and apparent distance modulus (0.12 and 17.47 mag) from the
web update of the Harris (1996) catalogue. The mass adopted for all stars is
0.85~M$_\odot$ and we used a bolometric magnitude of  $M_{\rm bol,\odot} = 4.75$
for the Sun. 

At variance with the analysis of previous clusters, in the case of Ter~8  we
used the strong Na D lines at 5889.97 and 5895.94~\AA\ in addition to the
strongest of the two lines at 5682-88~\AA, to have Na abundances based on
detections, rather than upper limits. Of course, these lines are available only
for stars observed with UVES. Furthermore, the abundances derived from strong
lines (exceeding 200~m\AA, like the Na D or the Ba~{\sc ii} lines) would result
in clear trends as a function of $v_t$, were the values of microturbulence
obtained with the usual method employing weaker iron lines. In the present work
we then decided to adopt the $v_t$ obtained from the relation as a function of
gravity used in Worley et al. (2013) for giants in
M~15, a metal-poor GC of similar metallicity.

Maintaining constant temperatures, gravities and microturbulent velocities,  
the abundances matching those derived from  Fe {\sc i} lines were interpolated 
within the Kurucz (1993) grid of model atmospheres (with the option for
overshooting on) to derive the final abundances. Note that  this choice has
a minimal impact on the derived abundances with respect to using models with no
overshooting. To check this we repeated the analysis of the six stars observed
with UVES using the option with no overshooting, and we found that the effect of
this change is negligible. Had we used models with no overshooting we would have
obtained [O/Fe] ratios larger by 0.033 dex on average, and [Na/Fe] ratios
smaller by 0.007 dex. The remaining element ratios would have changed less than
1 hundredth of dex in most cases.  The adopted atmospheric parameters and iron
abundances are listed in Table~\ref{t:atmpar08}  (only available in electronic
form) for individual stars.  

\setcounter{table}{3}
\begin{table*}
\centering
\caption[]{Sensitivities of abundance ratios to variations in the atmospheric
parameters and to errors in the equivalent widths, and errors in abundances for
stars of Ter~8 observed with UVES.}
\begin{tabular}{lrrrrrrrr}
\hline
Element     & Average  & T$_{\rm eff}$ & $\log g$ & [A/H]   & $v_t$    & EWs     & Total   & Total      \\
            & n. lines &      (K)      &  (dex)   & (dex)   &kms$^{-1}$& (dex)   &Internal & Systematic \\
\hline        
Variation&             &  50           &   0.20   &  0.10   &  0.10    &         &         &            \\
Internal &             &   5           &   0.04   &  0.08   &  0.10    & 0.10    &         &            \\
Systematic&            &  51           &   0.06   &  0.10   &  0.04    &         &         &            \\
\hline
$[$Fe/H$]${\sc  i}& 43 &    +0.087     & $-$0.025 &$-$0.022 & $-$0.022 & +0.016  &0.034    &0.095      \\
$[$Fe/H$]${\sc ii}& 12 &  $-$0.022     &   +0.067 &  +0.012 & $-$0.014 & +0.030  &0.037    &0.046      \\
$[$O/Fe$]${\sc  i}&  1 &  $-$0.062     &   +0.094 &  +0.040 &   +0.020 & +0.104  &0.112    &0.072      \\
$[$Na/Fe$]${\sc i}&  3 &  $-$0.021     & $-$0.041 &  +0.027 & $-$0.021 & +0.060  &0.068    &0.061      \\
$[$Mg/Fe$]${\sc i}&  2 &  $-$0.036     & $-$0.011 &  +0.001 &   +0.002 & +0.074  &0.074    &0.052      \\
$[$Al/Fe$]${\sc i}&  1 &  $-$0.046     &   +0.007 &  +0.008 &   +0.022 & +0.104  &0.107    &0.090      \\
$[$Si/Fe$]${\sc i}&  1 &  $-$0.067     &   +0.031 &  +0.017 &   +0.021 & +0.104  &0.107    &0.080      \\
$[$Ca/Fe$]${\sc i}& 16 &  $-$0.023     & $-$0.004 &  +0.001 &   +0.018 & +0.026  &0.032    &0.029      \\
$[$Sc/Fe$]${\sc ii}& 7 &    +0.028     &   +0.000 &  +0.004 &   +0.010 & +0.039  &0.041    &0.035      \\
$[$Ti/Fe$]${\sc i}& 16 &    +0.027     & $-$0.013 &$-$0.009 &   +0.005 & +0.026  &0.028    &0.036      \\
$[$Ti/Fe$]${\sc ii}& 6 &    +0.020     & $-$0.012 &$-$0.002 & $-$0.002 & +0.042  &0.043    &0.035      \\
$[$V/Fe$]${\sc i} &  3 &    +0.021     & $-$0.009 &$-$0.004 &   +0.020 & +0.060  &0.063    &0.033      \\
$[$Cr/Fe$]${\sc i}&  7 &    +0.011     & $-$0.013 &$-$0.007 &   +0.006 & +0.039  &0.040    &0.025      \\
$[$Cr/Fe$]${\sc ii}& 2 &  $-$0.005     & $-$0.011 &$-$0.008 &   +0.010 & +0.074  &0.075    &0.039      \\
$[$Mn/Fe$]${\sc i}&  4 &    +0.004     & $-$0.008 &$-$0.003 &   +0.016 & +0.052  &0.054    &0.023      \\
$[$Ni/Fe$]${\sc i}&  8 &  $-$0.014     &   +0.008 &  +0.005 &   +0.015 & +0.037  &0.040    &0.019      \\
$[$Cu/Fe$]${\sc i}&  1 &  $-$0.037     & $-$0.067 &  +0.004 &   +0.038 & +0.104  &0.112    &0.055      \\
$[$Zn/Fe$]${\sc i}&  1 &  $-$0.097     &   +0.055 &  +0.023 &   +0.015 & +0.104  &0.108    &0.108      \\
$[$Y/Fe$]${\sc ii}&  1 &    +0.122     & $+$0.033 &  +0.001 &   +0.038 & +0.104  &0.112    &0.132      \\
$[$Ba/Fe$]${\sc ii}& 3 &  $+$0.042     & $-$0.037 &$-$0.071 & $-$0.022 & +0.060  &0.086    &0.054      \\
$[$Nd/Fe$]${\sc ii}& 3 &    +0.052     & $+$0.052 &  +0.004 & $-$0.032 & +0.060  &0.069    &0.142      \\
\hline
\end{tabular}
\label{t:sensitivityu08}
\end{table*}

\begin{table*}
\centering
\caption[]{Sensitivities of abundance ratios to variations in the atmospheric
parameters and to errors in the equivalent widths, and errors in abundances for
stars of Ter~8 observed with GIRAFFE.}
\begin{tabular}{lrrrrrrrr}
\hline
Element     & Average  & T$_{\rm eff}$ & $\log g$ & [A/H]   & $v_t$    & EWs     & Total   & Total      \\
            & n. lines &      (K)      &  (dex)   & (dex)   &kms$^{-1}$& (dex)   &Internal & Systematic \\
\hline        
Variation&             &  50           &   0.20   &  0.10   &  0.10    &         &         &            \\
Internal &             &   5           &   0.04   &  0.12   &  0.10    & 0.14    &         &            \\
Systematic&            &  51           &   0.06   &  0.07   &  0.03    &         &         &            \\
\hline
$[$Fe/H$]${\sc  i}& 14 &    +0.065     & $-$0.014 &$-$0.011 & $-$0.012 & +0.038  &0.042    &0.074     \\
$[$Fe/H$]${\sc ii}&  1 &  $-$0.017     &   +0.075 &  +0.008 & $-$0.004 & +0.141  &0.142    &0.048     \\
$[$Na/Fe$]${\sc i}&  2 &  $-$0.039     & $-$0.019 &  +0.014 &   +0.010 & +0.100  &0.102    &0.096     \\
$[$Mg/Fe$]${\sc i}&  1 &  $-$0.032     &   +0.003 &  +0.003 &   +0.009 & +0.141  &0.141    &0.041     \\
$[$Si/Fe$]${\sc i}&  2 &  $-$0.045     &   +0.020 &  +0.007 &   +0.009 & +0.100  &0.101    &0.056     \\
$[$Ca/Fe$]${\sc i}&  4 &  $-$0.017     & $-$0.003 &$-$0.001 & $-$0.007 & +0.071  &0.071    &0.031     \\
$[$Sc/Fe$]${\sc ii}& 5 &  $-$0.051     &   +0.083 &  +0.022 &   +0.008 & +0.063  &0.071    &0.062     \\
$[$Ti/Fe$]${\sc i}&  2 &    +0.002     & $-$0.003 &$-$0.000 &   +0.010 & +0.100  &0.100    &0.037     \\
$[$V/Fe$]${\sc i} &  2 &    +0.012     & $-$0.004 &$-$0.000 &   +0.011 & +0.100  &0.100    &0.032     \\
$[$Cr/Fe$]${\sc i}&  1 &    +0.011     & $-$0.003 &$-$0.002 &   +0.015 & +0.141  &0.142    &0.047     \\
$[$Ni/Fe$]${\sc i}&  2 &    +0.003     &   +0.005 &  +0.002 &   +0.007 & +0.100  &0.100    &0.021     \\
$[$Ba/Fe$]${\sc ii}& 1 &  $-$0.041     &   +0.065 &  +0.007 & $-$0.069 & +0.141  &0.158    &0.099     \\
\hline
\end{tabular}
\label{t:sensitivitym08}
\end{table*}

The details of our procedure to derive errors in the atmospheric parameters are
given in Carretta et al. (2009a,b) for UVES and GIRAFFE respectively. To
evaluate the sensitivity of the derived abundances to the adopted atmospheric
parameters  we repeated our abundance analysis by changing only one atmospheric
parameter each time. The amount of the variations in the atmospheric parameters
and the resulting variations in abundances of Fe, O, Na, and all elements
measured  (i.e., the sensitivities) are shown in Table~\ref{t:sensitivityu08} and
Table~\ref{t:sensitivitym08}, for GIRAFFE and UVES spectra, respectively.  In the
upper part of the same tables we also give the error estimates for each
parameter. The typical internal error in $v_t$ was evaluated from the rms scatter of
the relation between $v_t$ and $\log g$ from Worley et al. (2013). The Column
labeled ``Total internal" gives the total star-to-star  error expected from
uncertainties in the atmospheric parameters and in the $EW$s.

Average abundances for iron and other elements in our sample are listed in
Table~\ref{t:meanabuTer8}, with those from the study by Mottini et
al. (2008) for comparison. Their abundances were corrected to our scale of
adopted solar reference abundances (from Gratton et al. 2003). 
From the analysis of the six stars with high-resolution UVES spectra the mean
metallicity we found for Ter~8
is  [Fe/H]$=-2.271 \pm0.032 \pm0.095$ dex ($rms=0.079$ dex); the
first error bar is from statistics and the second one refers to the systematic
effects. From the larger sample of 14 stars with GIRAFFE spectra we  obtained a
value of  [Fe/H]$=-2.249 \pm0.033 \pm0.074$ dex ($rms=0.123$ dex), in excellent
agreement with the estimate from UVES.
Compared to the scatter expected from uncertainties in atmospheric parameters
and $EW$s ($0.028\pm0.012$ dex for UVES and $0.040\pm0.011$ dex for GIRAFFE) we
conclude that in Ter~8 no intrinsic metallicity dispersion
statistically different from that predicted from error analysis is present.

The abundances of iron obtained from singly ionized species are in nice
agreement with those from neutral lines: [Fe/H]{\sc ii}$=-2.28$ ($rms=0.08$ dex,
six stars) from UVES and  [Fe/H]{\sc ii}$=-2.22$ ($rms=0.09$ dex, five stars) from
GIRAFFE.  This supports our adopted atmospheric parameters.

\begin{table}
\centering
\caption{Mean abundances for Ter~8.}
\setlength{\tabcolsep}{1.5mm}
\begin{tabular}{lrccrccrcccrc}
\hline
\hline
Element &n  &avg &$rms$ &n  &avg &$rms$&n  &avg &$rms$\\	    
        &\multicolumn{3}{c}{UVES} &\multicolumn{3}{c}{GIRAFFE} &\multicolumn{3}{c}{Mottini08}\\
\hline
$[$O/Fe$]${\sc i}    &6 &  +0.39 &0.05 &   &	    &	  &3 &  +0.52 &0.18\\
$[$Na/Fe$]${\sc i}   &6 &  +0.25 &0.13 &10 &  +0.18 &0.27 &  &        &    \\
$[$Mg/Fe$]${\sc i}   &6 &  +0.47 &0.09 &10 &  +0.46 &0.08 &3 &  +0.77 &0.21\\
$[$Al/Fe$]${\sc i}   &6 &$<$0.96 &0.19 &   &	    &	  &  &        &    \\
$[$Si/Fe$]${\sc i}   &6 &  +0.25 &0.10 & 8 &  +0.43 &0.09 &3 &  +0.49 &0.23\\
$[$Ca/Fe$]${\sc i}   &6 &  +0.19 &0.04 &14 &  +0.19 &0.10 &3 &  +0.40 &0.11\\
$[$Sc/Fe$]${\sc ii}  &6 &$-$0.12 &0.05 &13 &$-$0.01 &0.08 &  &        &    \\
$[$Ti/Fe$]${\sc i}   &6 &  +0.05 &0.06 &11 &  +0.11 &0.12 &3 &  +0.07 &0.06\\
$[$Ti/Fe$]${\sc ii}  &6 &  +0.12 &0.07 &   &	    &	  &3 &  +0.16 &0.08\\
$[$V/Fe$]${\sc i}    &6 &$-$0.30 &0.06 &10 &  +0.01 &0.09 &  &        &    \\
$[$Cr/Fe$]${\sc i}   &6 &$-$0.41 &0.06 & 3 &$-$0.12 &0.08 &1 &$-$0.25 &    \\
$[$Mn/Fe$]${\sc i}   &6 &$-$0.53 &0.05 &   &	    &	  &3 &$-$0.12 &0.12\\
$[$Fe/H$]${\sc i}    &6 &$-$2.27 &0.08 &14 &$-$2.25 &0.12 &3 &$-$2.40 &0.10\\
$[$Fe/H$]${\sc ii}   &6 &$-$2.27 &0.08 & 5 &$-$2.22 &0.09 &3 &$-$2.28 &0.14\\
$[$Ni/Fe$]${\sc i}   &6 &$-$0.18 &0.03 &14 &$-$0.01 &0.08 &3 &$-$0.09 &0.05\\
$[$Cu/Fe$]${\sc i}   &6 &$-$0.61 &0.08 &   &	    &	  &2 &$-$0.82 &0.09\\
$[$Zn/Fe$]${\sc i}   &6 &$-$0.05 &0.08 &   &	    &	  &  &        &    \\
$[$Y/Fe$]${\sc ii}   &6 &$+$0.09 &0.09 &   &	    &	  &3 &$-$0.35 &0.14\\
$[$Ba/Fe$]${\sc ii}  &6 &$-$0.15 &0.08 &14 &$-$0.29 &0.32 &3 &$-$0.15 &0.11\\
$[$Nd/Fe$]${\sc ii}  &6 &$-$0.30 &0.32 &   &	    &	  &2 &  +0.03 &0.18\\
$[$Eu/Fe$]${\sc ii}  & 6& $<$2.65 &    &   &	    &	  &3~& $<$1.45& ~~~~\\
\hline
\end{tabular}
\label{t:meanabuTer8}
\end{table}

\section{Results}\label{resu}

\subsection{Na and O abundances}\label{nao}

The main goal of our study was to determine the Na and O abundances to ascertain
whether also in this low-mass GC these elements show an intrinsic star-to-star
scatter and an anti-correlation, which is the main spectroscopic
signature of multiple populations in GCs.

Corrections for departures from the LTE assumptions to Na abundances  were
applied following Gratton et al. (1999). Abundances of oxygen were obtained from
$EW$s of the forbidden  [O~{\sc i}] 6300.31~\AA\ line measured on spectra
cleaned from telluric lines.

\begin{figure}
\centering
\includegraphics[bb=18 372 592 708, clip, scale=0.42]{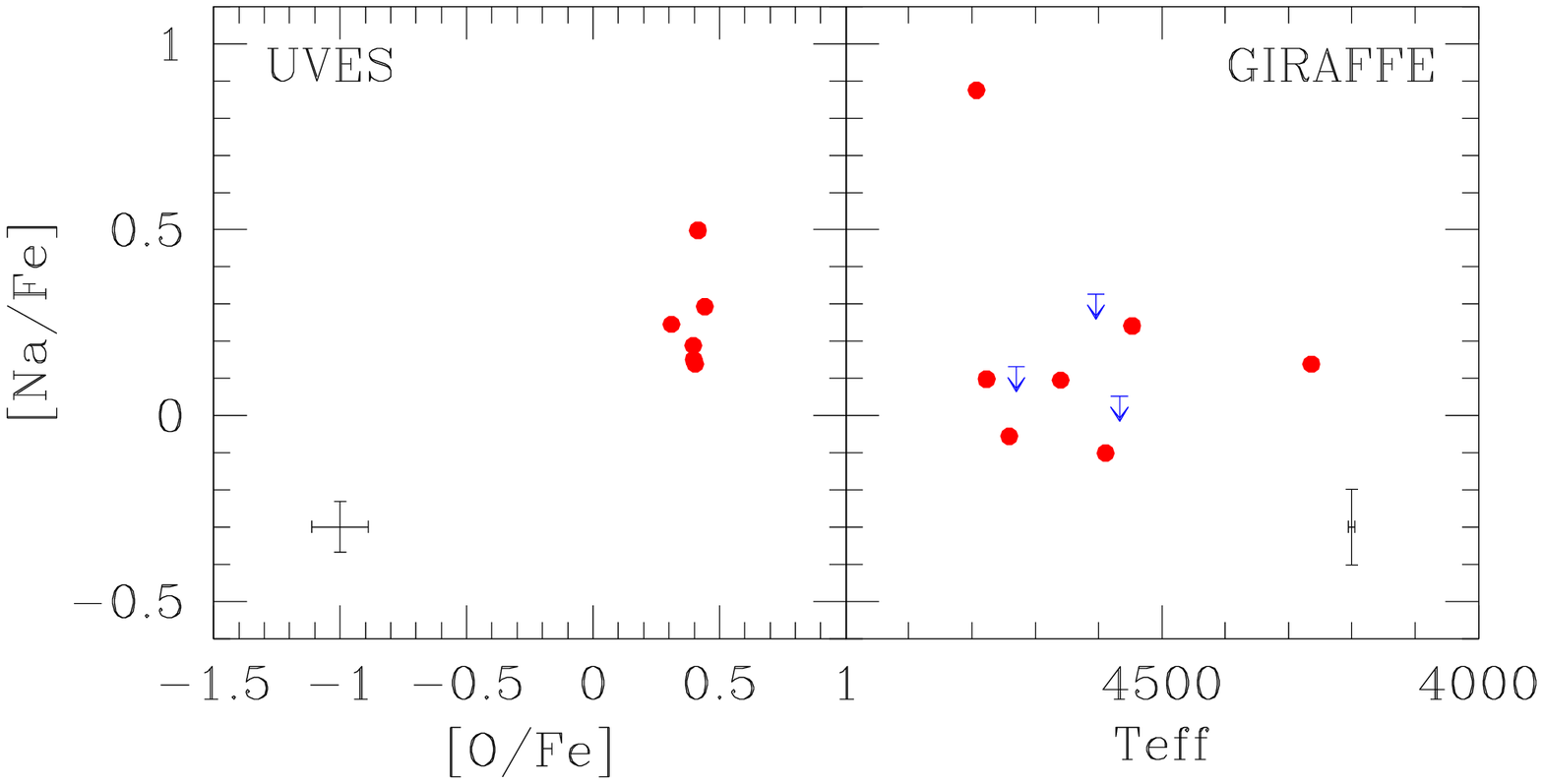}
\caption{Left: [Na/Fe] ratios as a function of [O/Fe] ratios for stars of Ter~8
with UVES spectra. Right: [Na/Fe] as a function of the effective temperature for
stars with GIRAFFE spectra. Upper limits in Na are indicated by arrows. In both
panels internal error bars are plotted.}
\label{f:nao08}
\end{figure}

In Fig.~\ref{f:nao08}, left panel, we plotted the results from UVES spectra for
stars in Ter~8. Apart from a small dispersion in Na, no anticorrelation between
Na and O abundances is discernible. From Table~\ref{t:sensitivityu08} and
Table~\ref{t:meanabuTer8} it is evident that no appreciable intrinsic scatter in
the [O/Fe] does exist among the six stars in Ter~8 observed with high-resolution UVES
spectra. We did not obtain the high average value for [O/Fe] derived by Mottini et al.
(2008) from their three stars, nor their large rms scatter (0.18 dex). We do not know the
cause of the discrepancy (which is, however, visible in most elements). The Solar values
for O are about the same (8.76 for them, 8.79 for us). The differences in
atmospheric parameters for the star 1188=305\footnote{The other star in common,
1209=325, was observed with GIRAFFE and we did not derive its oxygen abundance.
In this case, the atmospheric parameters are almost identical.} imply a
variation of less than +0.1 dex (i.e., in the wrong direction), using the
sensitivities of Table~\ref{t:sensitivityu08}.

The results we obtained from the GIRAFFE spectra are summarized in the right
panel of Fig.~\ref{f:nao08} as a function of the effective temperature, since no
measurement of O lines was possible on these spectra, too noisy in the  
[O~{\sc i}] region. For Na, we were able to derive seven actual detections and three
upper limits. 

Stars observed with GIRAFFE have a mean value of [Na/Fe]$=+0.18\pm 0.09$ with a
rms of 0.27 dex. Once the most Na-rich star is excluded from the average, the
mean [Na/Fe] ratio becomes  $0.10\pm 0.04$ dex, while the rms scatter decreases
to 0.13 dex,  compatible with the star-to-star error (0.10 dex) expected from
uncertainties in the abundance analysis.  The Na-rich star 2023 has a [Na/Fe]
ratio of 0.88 dex, i.e. more than 5$\sigma$ away from the average value defined
by the other 9 stars.

\begin{figure}
\centering
\includegraphics[scale=0.40]{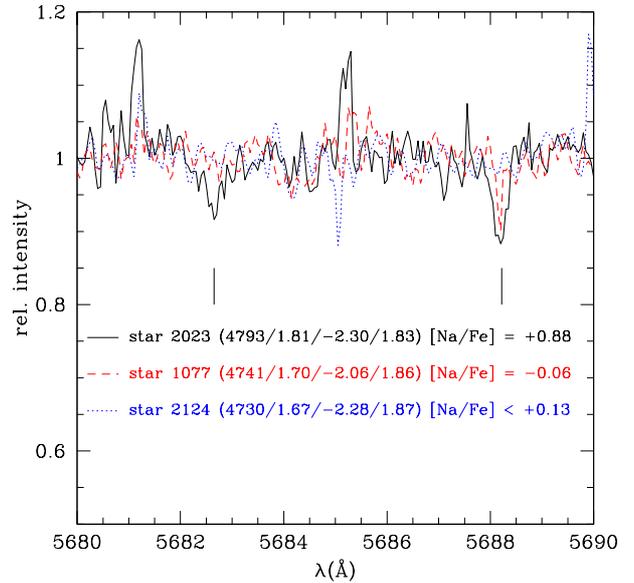}
\caption{Portion of the HR11 GIRAFFE spectrum for star 2023 in the region of the
Na doublet at 5682.65 and 5688.22~\AA \ (solid black line). The spectra of star
1077 (red dashed line) and of star 2124 (dotted blue line), with the most similar
atmospheric parameters in our sample, are superimposed. Sodium abundances for the
three stars are labelled.}
\label{f:m08p021}
\end{figure}

In Fig.~\ref{f:m08p021} we show the GIRAFFE spectrum of star 2023 in the region of
the Na~{\sc i} doublet at 5682-88 \AA, compared to the spectra of the two stars
with the most similar atmospheric parameters having Na measurements (one
detection and one upper limit).
All three objects are among the warmest in our sample, but, despite the not
excellent quality of the spectra, it is clear that star 2023 is different from
the comparison stars. Using Table~\ref{t:sensitivitym08}, it is possible to
evaluate that differences in atmospheric parameters should produce a cumulative
maximum difference in [Na/Fe] of less than 0.1 dex for both star 1077 and 2124.
The actual difference in the Na abundance with respect to star 2023 is much
larger than this, so we regard this as a robust detection of an high value of
[Na/Fe]. 

If the high Na content is intrinsic of this star, this object represents the
group of second-generation stars of Ter~8.  Alternatively, Na-rich material
could have been accreted from a companion star;  however, the RV of star 2023 is
in very good agreement with the cluster average, so it is  unlikely that it
belongs to a binary system or is a field interloper. It would be interesting to
further study this star, not an easy task, given its faint magnitude ($V=
16.98$). 

\subsection{Other proton-capture elements}

Other light elements typically involved in the network of proton-capture
reactions at high temperature are Mg, Al, and Si. For aluminum we only obtained
upper limits from the only transitions falling in the spectral range of UVES
spectra, the doublet Al~{\sc i} 6696-98~\AA. Abundances of these elements in
individual stars are listed in Table~\ref{t:proton08} and 
Table~\ref{t:alpha08} (only available in electronic form);
mean values obtained from UVES and GIRAFFE spectra are in
Table~\ref{t:meanabuTer8}.

\begin{figure}
\centering 
\includegraphics[scale=0.42]{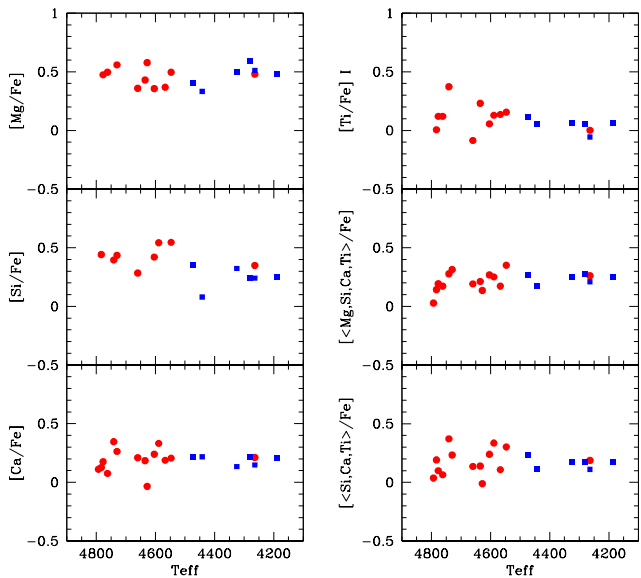}
\caption{Abundance ratios of $\alpha-$elements Mg, Si, Ca, Ti~{\sc i} as a
function of the effective temperature. The average of [$\alpha$/Fe] ratios are
shown in the last two panels on the right column (including and excluding the
Mg abundance from the mean, respectively). Red circles are for stars observed
with GIRAFFE, while blue squares indicate stars with UVES spectra. Internal 
star-to-star errors are listed in Table~\ref{t:sensitivitym08} and
Table~\ref{t:sensitivityu08}, respectively.}
\label{f:u08alpm08}
\end{figure}

The run of Mg and Si as a function of the effective temperature is plotted in
Fig.~\ref{f:u08alpm08} for stars with GIRAFFE (red circles) and UVES (blue
squares) spectra. Abundances of Mg and Si do not seem anticorrelated with
each other, especially when considering the associated error bars. This
occurrence points to a simple pre-enrichment by type II supernovae in the
proto-cluster cloud, that was not modified afterwards  by hot
proton-capture processes (see e.g. Yong et al. 2005, Carretta et al. 2009b for
examples of GCs with clear Si-Mg anticorrelation).

\subsection{Other elements}

Beside Mg and Si (potentially involved in proton-capture elements), we measured
the abundances of $\alpha-$elements Ca and Ti. Both these elements, in
particular Ca, do not present any intrinsic scatter, nor trend as a function of
the effective temperature (Fig.~\ref{f:u08alpm08}).

\begin{figure}
\centering 
\includegraphics[scale=0.42]{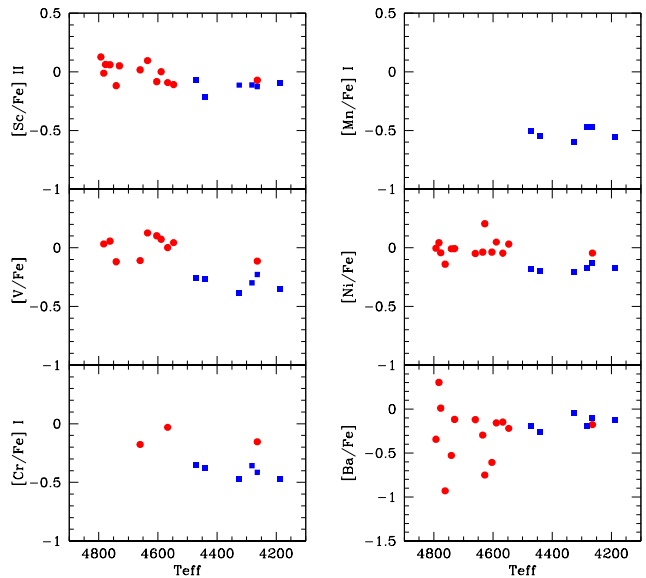}
\caption{As in Fig.~\ref{f:u08alpm08} for elements of the Fe-group (Sc~{\sc ii},
V, Cr~{\sc i}, Mn, Ni), and the $s-$process element Ba~{\sc ii}.}
\label{f:u08fegm08}
\end{figure}

Abundances for elements of the Fe-group Sc~{\sc ii}, V~{\sc i},
Cr~{\sc i}, and Ni~{\sc i} were obtained from both GIRAFFE and UVES spectra, and
are listed in Table~\ref{t:fegroup08}.
From Fig.~\ref{f:u08fegm08} there is good agreement between the
UVES and the GIRAFFE sample, apart from Cr and V, where an offset seems to be 
present between the two sets. From the UVES
spectra, with larger spectral coverage, we also derived abundances of 
Cr~{\sc ii}, Mn, and Zn. Details on the transitions used in the analysis can
be found in Carretta et al. (2011a); information on corrections due to the
hyperfine splitting adopted for Sc, V, and Mn are provided in Gratton et al.
(2003).

Abundances for Cu, Y, Ba, and Nd were derived following the procedure outlined in
Carretta et al. (2013). In the present case, due to the much lower metallicity
of the cluster under analysis, we could only place upper limits to the Eu
abundances.

\section{Discussion: stellar generations in Ter~8}

A summary of average abundances of several elements in GCs associated to the Sgr
dSph is given in Table~\ref{t:abugcsgr}. All these abundances are derived from
high resolution spectra. To these, we added the abundances derived for stars
belonging to the Sagittarius nucleus in Carretta et al. (2010c), as a
comparison. The abundance ratios for Pal~12, Ter~7 and Arp~2 are corrected to
our scale of solar reference abundances.

\setcounter{table}{9}
\begin{table*}
\centering
\caption{Mean abundances from high resolution spectra for GCs associated 
to Sgr.}
\begin{tabular}{lcccccc}
\hline
                     &   Ter 8       &   M~54        &     Sgr       &   Pal 12      & Ter 7         &Arp 2\\
Element              &  UVES         &    UVES       & UVES          & Cohen04       &Sbordone07     &Mottini08\\
                     &n~~~   avg~  $rms$ &n~~   avg~~  $rms$  &n~~   avg~~  $rms$ &n~~   avg~~  $rms$ & n~~   avg~~  $rms$ & n~~   avg~~  $rms$\\        
\hline
$[$O/Fe$]${\sc i}    &6   +0.39 0.05 &7 $-$0.02 0.47 &2 $-$0.00 0.09 &4   +0.02 0.11 &4   +0.15 0.06 &2   +0.10 0.04\\
$[$Na/Fe$]${\sc i}   &6   +0.25 0.13 &7   +0.33 0.39 &2 $-$0.19 0.27 &4 $-$0.37 0.04 &5 $-$0.16 0.10 &              \\
$[$Mg/Fe$]${\sc i}   &6   +0.47 0.09 &7   +0.28 0.09 &2   +0.09 0.06 &3   +0.25 0.01 &5   +0.08 0.07 &2   +0.53 0.18\\
$[$Al/Fe$]${\sc i}   &6 $<$0.96 0.19 &6   +0.65 0.62 &2 $-$0.09 0.33 &   	     &5 $-$0.11 0.13 &              \\
$[$Si/Fe$]${\sc i}   &6   +0.25 0.10 &7   +0.36 0.08 &2   +0.20 0.10 &4   +0.14 0.05 &5   +0.10 0.07 &2   +0.33 0.01\\
$[$Ca/Fe$]${\sc i}   &6   +0.19 0.04 &7   +0.32 0.08 &2   +0.14 0.01 &4 $-$0.08 0.04 &5   +0.12 0.13 &2   +0.46 0.01\\
$[$Sc/Fe$]${\sc ii}  &6 $-$0.12 0.05 &7 $-$0.08 0.12 &2 $-$0.22 0.02 &4 $-$0.10 0.04 &5 $-$0.26 0.10 &              \\
$[$Ti/Fe$]${\sc i}   &6   +0.05 0.06 &7   +0.18 0.10 &2   +0.03 0.08 &4 $-$0.09 0.03 &5   +0.10 0.07 &2   +0.17 0.08\\
$[$Ti/Fe$]${\sc ii}  &6   +0.12 0.07 &7   +0.27 0.12 &2   +0.08 0.21 &4 $-$0.04	0.05 &               &2   +0.17 0.08\\
$[$V/Fe$]${\sc i}    &6 $-$0.30 0.06 &7 $-$0.07 0.09 &2   +0.17 0.25 &4 $-$0.31 0.04 &5 $-$0.01 0.10 &              \\
$[$Cr/Fe$]${\sc i}   &6 $-$0.41 0.06 &7   +0.06 0.09 &2 $-$0.10 0.03 &4   +0.07 0.07 &5 $-$0.01 0.05 &2 $-$0.08 0.06\\
$[$Mn/Fe$]${\sc i}   &6 $-$0.53 0.05 &7 $-$0.49 0.09 &2   +0.01 0.14 &4 $-$0.21 0.03 &3 $-$0.21 0.07 &2 $-$0.29 0.05\\
$[$Fe/H$]${\sc i}    &6 $-$2.27 0.08 &7 $-$1.51 0.16 &2 $-$0.74 0.22 &4 $-$0.82 0.03 &5 $-$0.61 0.06 &2 $-$1.80 0.04\\
$[$Fe/H$]${\sc ii}   &6 $-$2.27 0.08 &7 $-$1.48 0.17 &2 $-$0.73 0.11 &4 $-$0.66 0.02 &5 $-$0.57 0.05 &2 $-$1.87 0.03\\
$[$Co/Fe$]${\sc i}   &               &7 $-$0.15 0.15 &2 $-$0.29 0.11 &4 $-$0.28 0.04 &5 $-$0.16 0.12 &2 $-$0.12 0.12\\
$[$Ni/Fe$]${\sc i}   &6 $-$0.18 0.03 &7 $-$0.09 0.03 &2 $-$0.17 0.07 &4 $-$0.19 0.05 &5 $-$0.20 0.05 &2 $-$0.07 0.14\\
$[$Cu/Fe$]${\sc i}   &6 $-$0.61 0.08 &7 $-$0.61 0.18 &2 $-$0.55 0.07 &4 $-$0.51 0.48 &3 $-$0.49 0.20 &2 $-$0.85 0.23\\
$[$Zn/Fe$]${\sc i}   &6 $-$0.05 0.08 &7   +0.03 0.15 &2 $-$0.14 0.02 &3 $-$0.48 0.13 &3 $-$0.27 0.20 &              \\
$[$Y/Fe$]${\sc ii}   &6   +0.09 0.09 &7 $-$0.18 0.17 &2   +0.09 0.25 &3 $-$0.51 0.13 &5 $-$0.18 0.19 &2 $-$0.16 0.08\\
$[$Zr/Fe$]${\sc i}   &               &6 $-$0.12 0.07 &2 $-$0.43 0.23 &4 $-$0.18 0.06 &               &              \\
$[$Zr/Fe$]${\sc ii}  &               &7 $-$0.09 0.12 &2 $-$0.08 0.01 &   	     &      	     &              \\
$[$Ba/Fe$]${\sc ii}  &6 $-$0.15 0.08 &7   +0.17 0.12 &2   +0.38 0.32 &4   +0.15 0.01 &4   +0.32 0.12 &2   +0.15 0.03\\
$[$La/Fe$]${\sc ii}  &               &7   +0.18 0.17 &2   +0.06 0.35 &4   +0.10 0.09 &5   +0.36 0.13 &2   +0.12 0.08\\
$[$Nd/Fe$]${\sc ii}  &5 $-$0.30 0.32 &7   +0.49 0.07 &2   +0.63 0.36 &3   +0.27 0.06 &3   +0.42 0.19 &2 $-$0.03 0.07\\
$[$Eu/Fe$]${\sc ii}  &6 $<$2.65~~~~~~~~ &7   +0.46 0.08 &2   +0.54 0.19 &4   +0.55 0.06 &               &2   +0.40 0.11\\
\hline
\end{tabular}
\begin{list}{}{}
\item[] References for the analyses are:\\
 M~54: Carretta et al. (2010b,c); 
 Sgr: Carretta et al. (2010b,c);
 Pal~12: Cohen (2004);
 Ter~7: Sbordone et al. (2007, we do not use Tautvaisiene et al. (2004) since their three stars are in this sample; the reader is directed to the Sbordone et al. paper for comparison between their results);
 Arp~2: Mottini et al. (2008).
\end{list}
\label{t:abugcsgr}
\end{table*}

\begin{figure}
\centering
\includegraphics[scale=0.45]{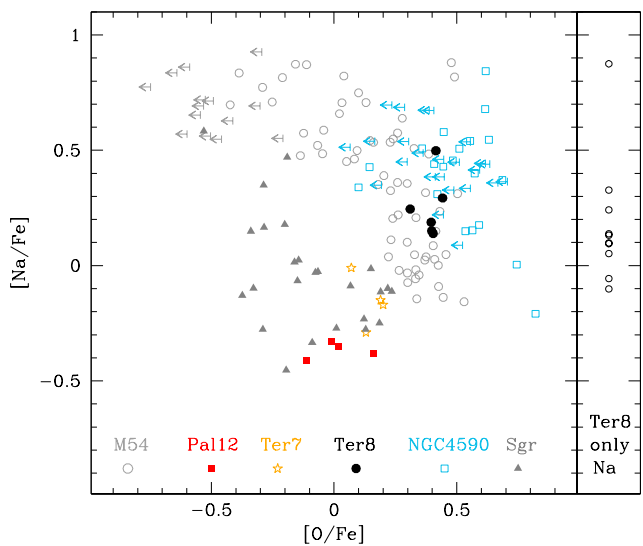}
\caption{The Na-O anticorrelation for individual stars in all confirmed Sgr GCs
(plus NGC~4590). For Ter~8, we also show the variation in Na for all the stars
with GIRAFFE spectra in the right panel. In the left panel, M~54 (grey empty
circles), NGC~4590 (empty squares), and Ter~8 (filled black circles) come from
our homogeneous analysis (Carretta et al. 2010c, 2009a,b, and this work,
respectively); the abundances of  Pal~12 (Cohen 2004, red filled squares) and
Ter~7  (Sbordone et al. 2007, open star symbols) have been corrected to our
scale of solar abundances (see Gratton et al. 2003). Filled triangles indicate
stars of the nucleus of Sgr from Carretta et al. (2010c).}
\label{f:naosgr}
\end{figure}

\subsection{Na, and O: Terzan 8 in context}

The Na-O anticorrelation is one of the most notable tracers of multiple stellar
populations in GCs (see Kraft 1994, Gratton et al. 2001, and Carretta et al.
2010a for summaries about its discovery, meaning, and distribution among GCs).
This chemical signature is so widespread among Galactic GCs that
can be assumed to be the distinctive signature of a genuine GC, as the chemical
relation established by the complex chain of events leading to the appearance of
a GC  (Carretta et al. 2010a). The  open clusters did not seem to present
the same chemical pattern, not even the most massive and old. Two of them have
been recently investigated in detail: Berkeley~39  and NGC~6791, of similar age
and mass. The first has a very homogeneous composition in all studied elements
(Bragaglia et al. 2012), the second shows a bimodal distribution in Na abundance
and a small spread in O that may be interpreted as a Na-O anticorrelation
(Geisler et al. 2012). We are analysing additional data to further 
investigate this interesting cluster.

While most of the MW GCs for which Na and O abundances have been  derived show
the Na-O anticorrelation (Carretta et al. 2010a), there seem to be a few
exceptions, like Pal~12 (Cohen 2004) and Ter~7 (Tautvaisiene et al. 2004,
Sbordone et al. 2005), two clusters with a very low present-day mass which do
not show any sign of this relation among the (admittedly small) samples of stars
investigated so far. Another case seems to be Rup~106, where Villanova et
al. (2013), found no significant variation in Na and O for the nine 
stars examined. It is more
massive than the other two, with a present-day mass larger than that of other
GCs showing instead a Na-O anticorrelation (see Fig. 1 in Bragaglia et al.
2012). It would be interesting to study this cluster with a sample larger than
the 9 stars available today, to reach more definitive conclusions on the
presence or absence of a secong generation of star. We note that Rup~106 has a
young age and is the only object among these three GCs that is not associated to
the Sgr dwarf galaxy.

The definition of {\it bona fide} globular cluster proposed by Carretta 
et al. (2010) uses a physical, measurable property of clusters (the presence 
of a peculiar chemical pattern), instead of rather fuzzy definitions based on
age (but not all GCs are old), or metallicity (not all GCs are metal-poor), or
position in the Galaxy (not all GCs reside in the halo), etc. It is a
work-in-progress definition, based on the observations and theoretical models
presently available. Cases like Rup~106 (not-so-small mass but no Na-O
anticorrelation, see Villanova et al. 2013) should be investigated to understand
the reason why this aggregate of stars avoided to develop different stellar
generations with distinct chemistry.

The Na, O abundances of individual stars in GCs found to be related to the Sgr
dSph are displayed in Fig.~\ref{f:naosgr}. Different symbols indicate different
GCs and, as a comparison, we also plot stars of the Sgr nucleus homogeneously
analysed by our group (Carretta et al. 2010c; filled grey triangles). Abundances
of stars in Pal~12 and Ter~7 were corrected to our scale of solar abundances
(Gratton et al. 2003) using the solar abundances adopted in the original papers
(Cohen 2004 and Sbordone et al. 2007, respectively).

From this figure we can appreciate how only the most massive GCs (M~54=NGC~6715)
and NGC~4590 (M68) show the presence of a Na-O anticorrelation among their
stars. In particular, M~54, sitting in the nucleus of Sgr, 
shows this feature both in its metal-poor and metal-rich components (not
separated in Fig.~\ref{f:naosgr} (see Carretta et al. 2010c). 
Though showing a difference of almost three 
orders of magnitude in mass (using the absolute magnitude as proxy of the
present-day mass), both M~54 and M~68 do participate to the well defined
relation that links the extension of the Na-O anticorrelation and present-day
mass (absolute magnitude), found by Carretta et al. (2010a) in their FLAMES
survey (see Carretta et al. 2013 for the most recent version). This evidence
suggests that, despite being formed not in the main Galaxy, but instead in one
of its dwarf satellites, the most massive GCs associated to Sgr were subject to
the same formation mechanism as ``normal" Galactic GCs. The presence of the Na-O
anticorrelation, although of different extension in the two objects, indicates
that a first generation must have been formed, evolved, and its most massive
members polluted the intracluster gas giving birth to one (or more, as likely 
in the case of M~54, Carretta et al. 2010c) further stellar generation.

The field component of the Sgr dwarf galaxy, represented in Fig.~\ref{f:naosgr}
by  stars of the nucleus (Carretta et al. 2010c), presents some dispersion in  
both O and Na. These two elements show no evidence of being anticorrelated with
each other in this component. Two stars of the nucleus  lie on the lower
envelope of the Na-O anticorrelation in M~54 and they could be tentatively
assumed as candidate cluster stars lost to the Sgr nucleus. The few stars
studied in Pal~12 (Cohen 2004) and Ter~7 (Sbordone et al. 2007) seem
to share the location of stars in the nucleus of Sgr (of which they share the
high metallicity)  in the Na-O plane. On average (see Table~\ref{t:abugcsgr},
where however only the mean from the two stars with high-resolution UVES spectra
are listed for the Sgr nucleus) these stars show a slight overabundance of O,
with Na about 0.2 dex subsolar. 
They occupy the typical location of Galactic field stars, with high-O
and low-Na abundances. Their position, at lower O values, than MW ones, is well explained by the
longer  times involved in the chemical evolution of dwarf spheroidals like Sgr:
the higher level of iron, due to the contribution from a relevant number of 
SN Ia in the enrichment of the intergalactic gas, decreases the [O/Fe] ratio
down to solar values (see however an alternative scenario in the very recent
paper by McWilliam et al. 2013, where the low $\alpha-$element content is
attributed to an initial mass function devoid of the highest mass stars). This explanation is valid also for Pal~12 and Ter~7.
Both these GCs are found to be younger than the bulk of the Galactic GCs
(see e.g. Carretta et al. 2010a) and also younger than the metal-poor GCs
in Sgr (M~54 and M~68). It is likely that stars in Pal~12 and Ter~7 formed from
gas already experiencing higher enrichment from type Ia SNe.

On the other hand, the six giants with UVES spectra in Ter~8 have O and Na
abundances compatible with those of M~54 and M~68, and about 0.5 dex higher
than the average values observed in Pal~12 and Ter~7. One giant in Ter~8
(not shown in Fig.~\ref{f:naosgr} since it has no O measured) shows a
high [Na/Fe] ratio, compatible with a composition modified by proton capture
reactions (intermediate component in the classification scheme proposed by
Carretta et al. 2009a). If the chemical evolution in Ter~8 was on the verge to 
develop a Na-O anticorrelation, this  may have some impact on the theories
illustrating the GC formation. Ter~8 has a present-day mass very  similar to the
one of Ter~7, but the second is more metal-rich. The implication is that in
addition to the total mass also the metal abundance is a crucial factor: at the
same mass, a lower metallicity is favoured for building up  a second generation
within a cluster. In the best (closest) example of dwarf spheroidal galaxy with
its own system of GCs available to us, two stellar generations seem to have
appeared only amongst the most metal-poor objects.

\subsection{Is Ter~8 a FG-only or mainly-FG cluster?}

Caloi and D'Antona (2011) presented a list of candidate GCs only (or mainly) 
composed by FG stars (Ter~8 and Arp~2 are among them). They based their
selection on the possibility of producing the observed HBs with (almost) a
single mass value. This occurrence would indicate the absence of any 
significant spread in He, and therefore of SG stars. The difference between  the
two proposed categories is that FG-only GCs were not massive enough to retain
the gas necessary to form a SG, while  mainly-FG clusters could form a SG but did
not lose the vast majority of their  primordial stars (SG stars are supposed to
be currently the majority only because most of the FG was lost by the clusters,
see e.g., D'Ercole et al. 2008).

In the present paper we found that there are some Na variations in Ter~8, but
the fraction of Na-rich stars is much lower than for most of the GCs studied to
date.  This is evidence for stars with mostly primordial composition. Only a few
objects may be part of a SG, the opposite of what we find for more massive GCs. 
The binomial distribution is the best suited tool to approach statistically the
issue, since one star could belong either to the first- or to the
second-generation. Using this distribution, finding only one SG star over 16
analysed means that  we can exclude that SG stars are more than 30\% in this
cluster  with  a 98 percent probability, and that they are more than 15\% with a
95 percent probability.

Considering together the three low-mass Sgr GCs (Pal~12, Ter~7, and Ter~8), we
have one bona fide SG star over 25 (zero over four and five, respectively, in
Ter~7  and Pal~12). Using again the binomial distribution, this  would imply
that the probability that SG stars are more than 30\% is tiny (about 0.1
percent).  
Can we conclude that all the Sgr GCs are examples of 
FG-only or mainly-FG clusters? As discussed above, the obvious counter-example 
is M~54, which has one of the more extended Na-O anticorrelation found so far 
(Carretta et al. 2010b,c). However, this is a very massive GC. Another possible
exception is NGC~4590 (M~68; see the previous subsection), although we remind
the reader that its attribution to the Sgr system of GCs is still disputed (see e.g.,
Dalessandro et al. 2012). The observed Na-O anticorrelation in this GC is not
very extended and this cluster shows one of the highest fraction of FG stars in
our FLAMES survey, about 40\%.

We can also look at clusters in other dwarf galaxies.  The only two cases where
detailed abundance analysis of individual giant stars were measured are the
dwarf spheroidal Fornax (Fnx) and the dwarf irregular Large  Magellanic Cloud
(LMC). Letarte et al. (2006) measured nine stars in three  GCs of Fnx, Fornax
1, 2, and 3, with total absolute magnitudes $M_V=-5.32$,  $-7.03$, and $-7.66$,
respectively (van den Bergh \& Mackey 2004).  These GCs are all very metal-poor,
with [Fe/H] about -2.3 (Letarte et al. 2006). Only two of the nine analysed
stars are classified SG. One star is in Fornax~1, whose $M_V$ is similar to the
one of Ter~8 ($M_V=-5.07$ in the Harris 1996 catalogue and $-5.68$ in Salinas et
al. 2012). The second star is in Fornax~3,  which has a much larger mass. Taken
at face value, this means a fraction of about 20\% of SG. On the other hand,
D'Antona et al. (2013) analysed the HB of the  Fornax clusters. They found no
evidence for a second generation in Fornax 1  ($M_V=-5.2$, Webbink 1985) while
according to their analysis the remaining brighter  clusters ($-8.2<M_V<-7.2$)
should contain half or more second-generation stars.  The real fraction of
second generation stars in these clusters is then controversial.

Moving to the LMC, Mucciarelli et al. (2009) found a Na-O anticorrelation in
three metal-poor ([Fe/H] $\sim -1.8$), massive ($M_V=-7.25, ~-7.51, ~-7.70$, van
den Bergh \& Mackey 2004), old clusters. In this case, however, the proportion
between FG and SG among the 18 stars analysed is the same as for the massive MW
GCs. Maybe this occurrence is due to the different kind of galaxy (dwarf
irregular versus spheroidal) but we do not have enough data to draw firm
conclusions. No spread in Na and O was found in younger and slightly less
massive LMC clusters (Mucciarelli et al. 2008). On the other hand, Johnson,
Ivans \& Stetson (2006) found no significant variation in Na and O  among 10
stars they analysed in four old LMC GCs, even if they were rather massive ($M_V$
from $-7.40$ to $-7.75$, van den Bergh \& Mackey 2004). Johnson et al. however
found a large enhancement in Al in one of the stars, while Hill et al. (2000)
measuread a large Al spread in three stars of another LMC cluster. This evidence
is rather puzzling, because the temperatures required for the efficient action
of the Mg-Al cycle are much higher than those necessary for the activation of
the CNO and Ne-Na cycles. The situation for the LMC  clusters is still unclear
and needs to be further studied using larger and more homogeneous samples.

In N-body simulations built to represent scenarios of GC formation where the
main polluters are intermediate mass AGB stars (D'Ercole et al. 2008) or FRMS
(Decressin et al. 2010), a common feature is that the final product (the
currently observed cluster) should be more compact than its progenitor
(Vesperini et al. 2013). This naturally follows either from the cooling flow
collecting gas for the SG in the cluster centre (D'Ercole et al. scenario) or
from the birth or migration of very massive stars to the centre (Decressin et
al. scenario). Coupled to the other characteristics of FG stars, this prediction
has some consequences for GCs expected to be almost entirely composed or
dominated by FG stars. Among them we should expect: i) no spread at all in Na
and O (or a very small  one), ii) no large spread in He, and therefore a shorter
HB compared to GCs of similar mass and metallicity, iii) a lower concentration.
In turn, the last property implies a larger fraction of primordial binaries and
a population of blue stragglers and millisecond pulsars more similar to the one
found in the Galactic field.

Do GCs belonging to dSph galaxies share these peculiarities? As discussed above,
even when massive enough, GCs in Sgr and Fnx show a tendency to homogeneous
values of Na and O, at least for the majority of their stellar population.  The
situation is less clear for LMC clusters.
 
The dominance of FG over SG implies a smaller average He content
and a smaller dispersion of He abundance. This shows up very well on the HB,
whose extension is related to metallicity, age, and He (see e.g. Gratton et al.
2010 and references therein). Gratton et al. determined the dispersion in mass 
(i.e., in He, since higher He means lower mass on the HB) in a large number  of
GCs from their HBs. Unfortunately, only three of the GCs which are
probable members of Sgr  were studied: NGC~4590, NGC~5053, and NGC~5466. All
three GCs have a very low metallicity ([Fe/H]$\sim -2.3$ dex) and all seem to 
present a smaller spread in mass for their metallicity, lying clearly below the
best fit line in the mass loss vs [Fe/H] plane.
In the work on UV properties of GCs observed with the GALEX satellite 
(Dalessandro et al. 2012) the four most metal-poor GCs associated to Sgr (the
three mentioned above and Ter~8) are clearly seen separated from the bulk of
MW GCs in the ($FUV-V$) vs [Fe/H] plane, with Sgr clusters having
systematically redder colours than their Galactic counterparts of similar
metallicity. Again, this difference can be consistently explained by the 
different distribution of stars along the HB for MW and Sgr clusters of similar
metallicity and age. Dalessandro et al. found that the average R'-parameter
(relating the number of HB and RGB stars) as measured by Gratton et al. (2010)
is smaller for the Sgr clusters than for the MW ones, possibly implying a lower
He abundance. This points again towards a dominance of FG stars, since the SG
tend to have higher He, and, as a result, brighter and bluer HBs.

The lower concentration predicted for GCs predominantly composed by FG stars
seems well matched by the clusters probably belonging to Sgr, which tend to show
a lower concentration (see the Harris 1996 catalogue).

\begin{figure}
\centering
\includegraphics[scale=0.45]{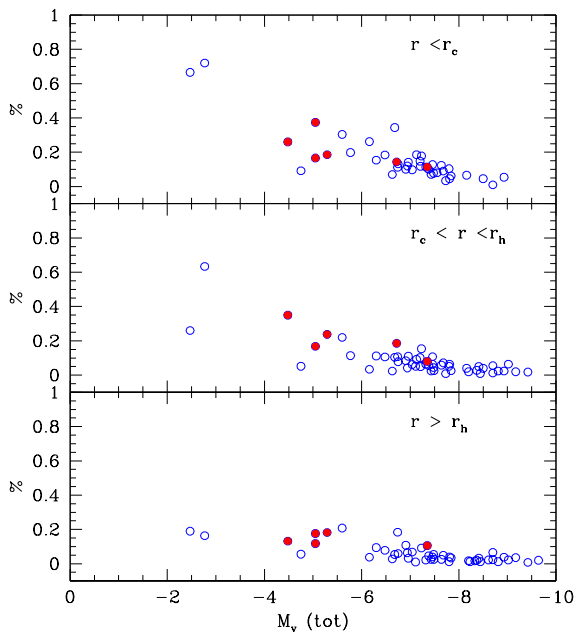}
\caption{Total fraction of binaries measured in GCs by Milone et al. (2012b) for
their $r_c$, $r_c$-HM (where HM indicates the half-mass radius $r_h$), and oHM
samples (from top to bottom, respectively; see the original paper for
definitions) as a function of the total absolute $V$ magnitude. Filled red
symbols indicate the GCs associated to the Sgr dSph (see text).}
\label{f:binfrac}
\end{figure}

Finally, a larger binary fraction should be found among FG stars than in SG 
ones, as we proposed in D'Orazi et al. (2010). This again stems from the 
different concentration of the two populations, the SG binaries being more
susceptible to be destroyed by collisions.  We used the binary fractions
measured by Milone et al. (2012b) and found that Arp~2, NGC~4590, NGC~5053,
Pal~12, Ter~7, and Ter~8 seem to have binary fractions larger than  the
majority of MW GCs, from more than $10$ up to 20\%, compared to $<5\%$ (see
Fig.~\ref{f:binfrac}, where three different regions are considered, from Milone
et al. 2012b). Admittedly, this is more a hint that a firm conclusion, given the
spread of the relations and the fact that lower mass GCs tend to have larger
binary fractions. However, among GCs of similar mass, the clusters associated to
the Sgr galaxy seem to show a tendency for higher binary fractions (especially
looking at the lower panel of the figure).

In summary, it is possible to identify some common features in GCs formed in
and/or presently residing in dwarf galaxies (especially in dwarf spheroidals).
They tend to be less concentrated and more dominated by FG stars, at variance
with those of the inner halo of the MW. To these signatures we can also add that
the escape velocity is usually low in these objects.

An interesting possible example of such a process caught in the act is the
double cluster NGC~1850a+b in the LMC (Gilmozzi et al. 1994). NGC~1850b, the younger component,
should correspond to the SG, in the framework depicted above, and is much
smaller  (about 10$^3$ M$_\odot$) compared to the mass 10$^5$ M$_\odot$ of
NGC~1850a, that should represent the older, FG component. Unless there is an
unknown mechanism producing selective mass loss from the latter, the FG will
dominate even after the merging of the two components.

\subsection{$\alpha-$ and heavier elements}

\begin{figure*}
\centering
\includegraphics[scale=0.65]{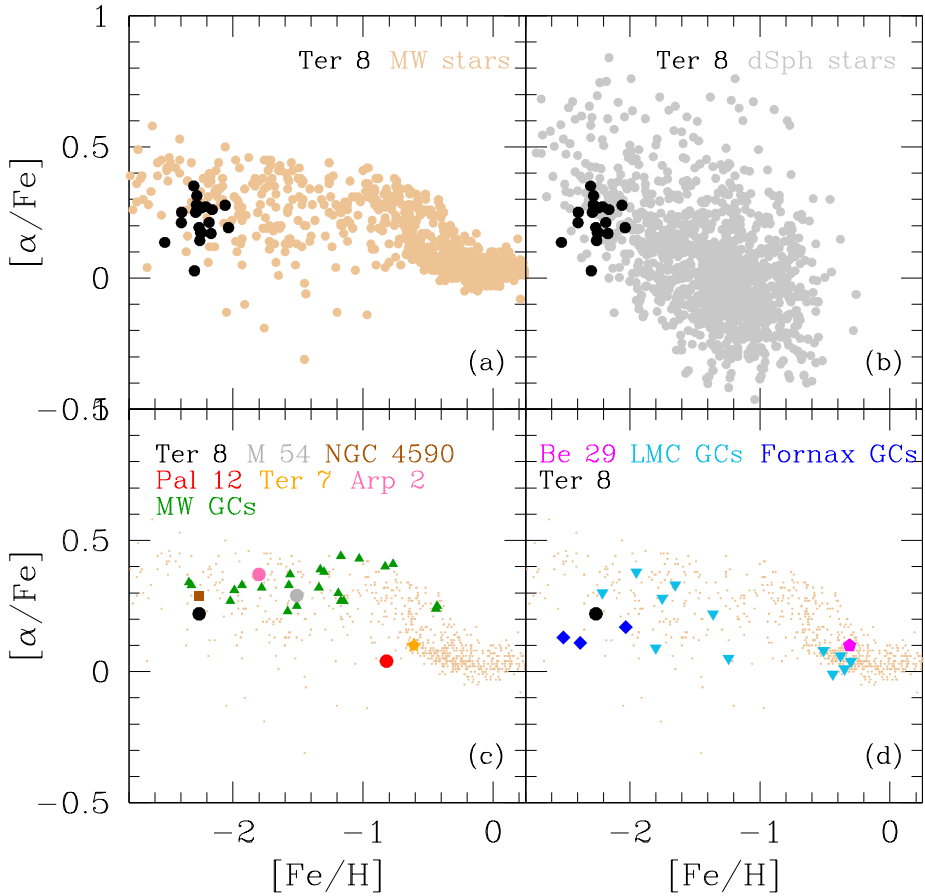}
\caption{Ratio [$\alpha$/Fe] versus metallicity in several Galactic and
extragalactic stellar populations. In (a) the ratios for individual stars
analysed in Ter~8 (large black circles) are superimposed to stars in our Galaxy
from the compilation by Venn et al. (2004; brown circles). In (b) the individual
values for Ter~8 are superimposed to those for stars in eight dwarf galaxies
from the extensive study by Kirby et al. (2010; grey points). The average
[$\alpha$/Fe] ratio for Ter~8 (black circle) is compared in panel (c) to the
average ratios of GCs associated to Sgr (M~54, grey circle; NGC~4590, brown
square; Arp~2, magenta circle; Ter~7, orange star symbol; Pal~12, red circle)
and to the mean ratios for MW GCs (green triangles). Finally, in 
(d) the average value for Ter~8 is compared to several extragalactic clusters:
GCs in the Fornax dwarf galaxy (blue diamonds), clusters in LMC (cyan
triangles), and Be~29, an open cluster associated to Sgr. As a comparison, the 
small brown dots in panels (c) and (d) indicate the field stars in the Galaxy 
from Venn et al. (2004).} 
\label{f:alfa}
\end{figure*}

The pattern of abundances of $\alpha-$elements found in stars of Ter~8 bears the
typical signature of nucleosynthesis from massive stars only. In panels (a)
and (b) of Fig.~\ref{f:alfa}, values of the [$\alpha$/Fe] ratio (the average of
Mg, Si, Ca, and Ti~{\sc i} abundances) for individual stars in Ter~8 are
superimposed to the values for Galactic field stars (from the compilation by
Venn et al. 2004) and giants in eight dSph galaxies (Kirby et al.
2010). Our sample is compatible with the overabundance of $\alpha-$elements also
shown by the comparison samples of the MW and dSph stars at similar
metallicities, suggesting a not significant contribution from type Ia SNe to the
nucleosynthesis.

The average [$\alpha$/Fe] ratio for Ter~8 is compared in panel (c) of
Fig.~\ref{f:alfa} to the average values for other GCs in the Galaxy, and in
panel (d) to extragalactic GCs. The mean abundance in Ter~8 well agrees with
that of other Galactic GCs, as well as with that of metal-poor GCs associated to
the Sgr dSph (NGC~4590, M~54, Arp~2). On the other hand, the level of
$\alpha-$elements in stars of the more metal-rich GCs of Sgr (Pal~12 and
Ter~7) is distinctly lower, and well evident if compared to the Galactic field
stars (Venn et al. 2004, ad references therein).

\begin{figure}
\centering
\includegraphics[scale=0.45]{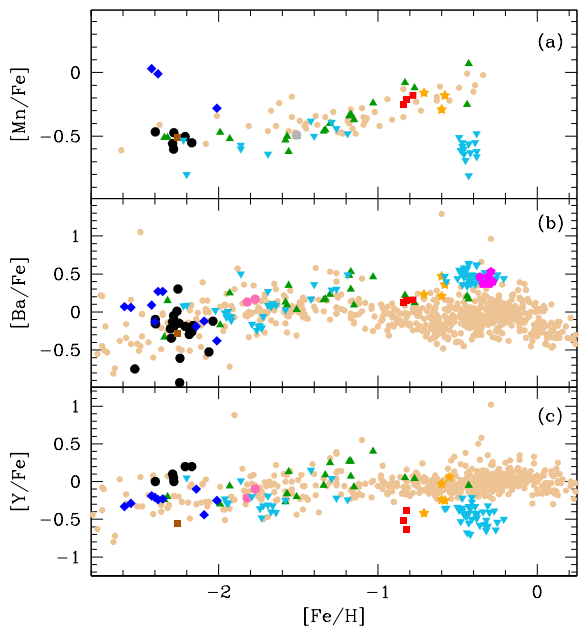}
\caption{As in the previous figure, but for the heavier elements Mn, Ba, and Y.
In this case we give average values for the MW GCs and individual abundance 
rations for the other clusters. Abundances of Mn for Galactic field stars are
from Gratton et al. (2003).}
\label{f:heavy}
\end{figure}

The abundance of Mn (an element related to the neutron excess and, as a
consequence, to the metal abundance) in Ter~8 perfectly agrees with those of
Galactic field stars of similar metallicities (Fig.~\ref{f:heavy}, panel (a)). 
The trend of [Mn/Fe] as a function of metal abundance displayed by field stars
in the Galaxy is rather well followed also by GCs in the MW and in Sgr, and is
mirroring the pattern of the $\alpha-$elements.

\begin{figure}
\centering
\includegraphics[scale=0.45]{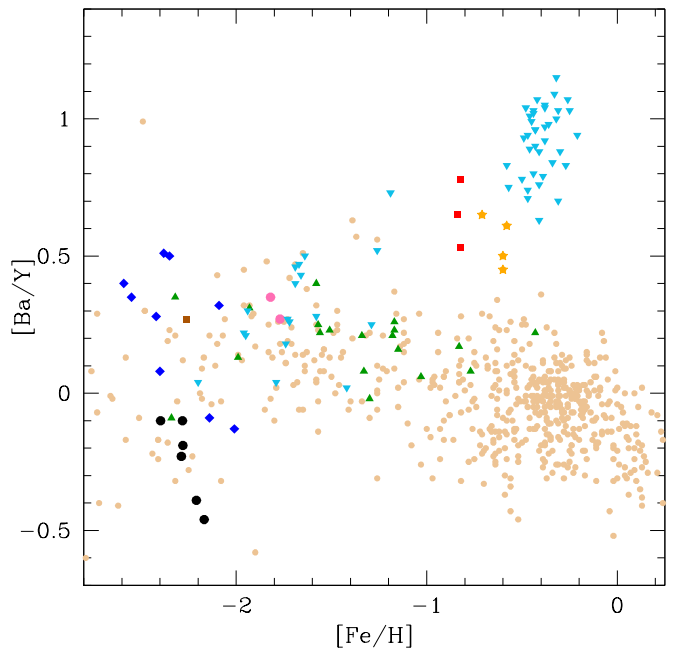}
\caption{Abundance ratios [Ba/Y] as a function of the metallicity. Symbols are
as in the previous Figure.}
\label{f:bay}
\end{figure}

Abundances of the light neutron-capture element yttrium and of the heavy 
neutron-capture element barium are plotted in panels (b) and (c) of
Fig.~\ref{f:heavy}, and their ratio is plotted in Fig.~\ref{f:bay}, as a
function of the metallicity. This ratio is important because Y and Ba sample
different peaks in the production of $s-$process elements; Y belongs to the 
first peak around neutron magic number N=50, whereas Ba belongs to the second 
peak that is built around N=82. The ratio [$hs$/$ls$] of heavy-to-light
$s-$process element is useful to probe the efficiency of the neutron-capture
process. For low efficiency, the neutron flux mainly feeds the first peak
nuclei, whereas the species in the second peak (like Ba) are favoured in case of
higher neutron exposures (higher efficiency, see Busso et al. 2001).
The Ba lines are not the best indicators, because they are generally 
strong, with limited sensitivities to changes in the abundances. A better
candidate would be La, that has similar nucleosynthetic history and $s-$process
contribution to the abundances in the solar system. Unfortunately, the La lines
are too weak to be detected in our spectra of  Ter~8.

The [Ba/Y] ratio of the most metal-rich GCs associated to Sgr (Pal~12 and Ter~7)
is large, about 0.5 dex, much higher than the bulk of Galactic field stars of
similar metallicity. This is true also for the
metal-rich Sgr stars, see Sbordone et al. (2007) and Smecker-Hane and McWilliam (2002). 
As discussed in the latter, the ratio [La/Y]=0.45 dex 
 could indicate a significant contribution of metal-poor AGB
stars: at low metallicity, the efficiency of the third dredge-up increases, as
well as the number of available neutrons, simultaneosly to the decrease of seed
nuclei. These occurrences shift the bulk of $s-$process production toward
heavier elements, in the second peak, so that the ratio [Ba/Y] (or [La/Y]) is
high. Our data in Ter~8 do not present such an evidence, although the sample is
too small to draw any firm conclusion.

\section{Summary}

We have observed  Ter~8, belonging to the Sgr dSph family, using FLAMES@VLT with
the intent of detecting a Na-O anticorrelation in this low-mass GC. From the
spectra of six stars observed with UVES and 10 with GIRAFFE, we have measured
Fe, Na, and O abundances (the last only in the UVES spectra). We did not detect
any O underabundance, and we found only one star that shows an
enhanced value of Na. 

In Ter~8 we found one star with the typical chemical composition of SG stars. In
this case, these stars represent a minority fraction, i.e. about 6\% of the
population, the opposite of what is
found for higher mass MW clusters (Carretta et al. 2010), and also at variance
with other low-mass Sgr GCs, Pal~12 a and Ter~7, where no significant Na and O
spread was seen. Ter~8 may represent a candidate for the class of  mainly-FG
cluster, while the two other GCs can be considered as good candidates for
FG-only clusters (Caloi \& D'Antona 2011), with the caveat that only a very few
stars were analysed in the last two clusters.

Not all Sgr dSph GCs studied present the same situation; a dichotomy is present
between the metal-poor, (very) high-mass M54 -and maybe NGC~4590- and the 
low-mass GCs. We
compared the case of Ter~8 to the clusters in Fornax and LMC showing variations
in Na and O. The environment where the clusters formed (i.e., a dwarf galaxy)
combines with cluster mass to define whether they could retain a large fraction
of the original population, hence showing a dominance of FG over SG. Further
studies of low-mass clusters in the MW and in neighbouring galaxies are required
to disentangle the effects and better constrain cluster formation mechanisms. In
particular, it would be interesting to target clusters which have a higher
probability of having been formed in dwarf galaxies and later accreted by the
MW. They should preferentially reside in the outer halo, could be younger, maybe
display peculiarities in their variables and generally should show the features
discussed above.

\begin{acknowledgements}
VD is an ARC Super Science Fellow. This publication makes use of data products
from the Two Micron All Sky Survey, which is a joint project of the University
of Massachusetts and the Infrared Processing and Analysis Center/California
Institute of Technology, funded by the National Aeronautics and Space
Administration and the National Science Foundation.  This research has been
funded by PRIN INAF 2011 "Multiple populations in globular clusters: their role
in the Galaxy assembly" (PI E. Carretta), and PRIN MIUR 2010-2011, project ``The
Chemical and Dynamical Evolution of the Milky Way and Local Group Galaxies'' (PI
F. Matteucci). We made use of the package CataPack, for which we are grateful to
Paolo Montegriffo. We thank Emanuele Dalessandro for his help with the GALEX data. 
This research has made use of the WEBDA, the SIMBAD database, operated at
CDS, Strasbourg, France, and of NASA's Astrophysical Data System.
\end{acknowledgements}

\clearpage

\Online
\begin{table*}
\centering \tiny
\setcounter{table}{1}
\caption{Information on the stars observed in Ter~8. The
complete table is available electronically only at CDS.}
\begin{tabular}{rcccccrrl}
\hline
\hline
 ID   &   RA       & Dec         & V     & I      & K	  & RV    &err  &Notes\\
      &(hh mm ss)  &(dd pp ss)   &       &        &(2MASS)&\multicolumn{2}{c}{km/s} &\\

\hline
 2357 &19 42 00.02 &-33 58 21.52 &15.072 &13.570  &11.697 & 144.64 &0.52 &UVES \\
 1658 &19 41 49.77 &-33 59 44.53 &15.269 &13.838  &12.033 & 146.07 &0.39 &UVES \\
  530 &19 41 55.64 &-34 01 48.12 &15.288 &13.881  &12.114 & 144.17 &0.53 &UVES \\
 1188 &19 41 37.06 &-34 00 33.78 &15.411 &14.069  &12.304 & 145.65 &0.66 &UVES \\
 3014 &19 41 23.69 &-33 56 03.30 &15.447 &14.082  &12.218 &   0.75 &0.52 &UVES,NM\\

\hline
\end{tabular}
\label{t:tabphot}
\end{table*}

\clearpage
\setcounter{table}{2}
\begin{table*}
\centering
\caption[]{Adopted atmospheric parameters and derived iron abundances.}
\begin{tabular}{rccccrcccrccc}
\hline
Star   &  $T_{\rm eff}$ & $\log$ $g$ & [A/H]  &$v_t$	     & nr & [Fe/H]{\sc i} & $rms$ & nr & [Fe/H{\sc ii} & $rms$ \\
       &     (K)	&  (dex)     & (dex)  &(km s$^{-1}$) &    & (dex)	  &	  &    & (dex)         &       \\
\hline
  1209 & 4264 & 0.80 & $-$2.16 & 2.14 & 20 & $-$2.160 & 0.092 &  1 &  $-$2.066 &       \\ 
  1169 & 4567 & 1.38 & $-$2.25 & 1.96 & 17 & $-$2.248 & 0.168 &    &	       &       \\ 
  2086 & 4547 & 1.33 & $-$2.30 & 1.98 & 13 & $-$2.301 & 0.147 &  2 &  $-$2.218 & 0.292 \\ 
   526 & 4589 & 1.42 & $-$2.40 & 1.95 & 13 & $-$2.395 & 0.128 &  1 &  $-$2.328 &       \\ 
  1728 & 4604 & 1.45 & $-$2.24 & 1.94 & 17 & $-$2.242 & 0.120 &    &	       &       \\ 
  2913 & 4628 & 1.49 & $-$2.52 & 1.93 & 10 & $-$2.525 & 0.067 &    &	       &       \\ 
   514 & 4635 & 1.50 & $-$2.19 & 1.93 & 17 & $-$2.184 & 0.139 &  1 &  $-$2.240 &       \\ 
   496 & 4660 & 1.54 & $-$2.04 & 1.91 & 17 & $-$2.035 & 0.158 &    &	       &       \\ 
   126 & 4777 & 1.79 & $-$2.26 & 1.84 & 13 & $-$2.261 & 0.134 &    &	       &       \\ 
  2124 & 4730 & 1.67 & $-$2.28 & 1.87 & 18 & $-$2.278 & 0.256 &  1 &  $-$2.271 &       \\ 
  1077 & 4741 & 1.70 & $-$2.06 & 1.86 & 11 & $-$2.060 & 0.071 &    &	       &       \\ 
  2531 & 4762 & 1.74 & $-$2.24 & 1.85 & 12 & $-$2.244 & 0.151 &    &	       &       \\ 
  2180 & 4783 & 1.79 & $-$2.25 & 1.84 & 14 & $-$2.255 & 0.224 &    &	       &       \\ 
  2023 & 4793 & 1.81 & $-$2.30 & 1.83 &  8 & $-$2.296 & 0.115 &    &	       &       \\ 
  2357 & 4188 & 0.66 & $-$2.29 & 2.18 & 52 & $-$2.288 & 0.100 & 13 &  $-$2.302 & 0.157 \\ 
  1658 & 4264 & 0.80 & $-$2.39 & 2.14 & 28 & $-$2.397 & 0.073 & 11 &  $-$2.400 & 0.132 \\ 
   530 & 4282 & 0.84 & $-$2.28 & 2.13 & 50 & $-$2.280 & 0.108 & 13 &  $-$2.293 & 0.092 \\ 
  1188 & 4325 & 0.92 & $-$2.28 & 2.10 & 32 & $-$2.282 & 0.112 & 11 &  $-$2.264 & 0.117 \\ 
  2253 & 4442 & 1.14 & $-$2.17 & 2.04 & 51 & $-$2.168 & 0.117 & 14 &  $-$2.152 & 0.128 \\ 
   137 & 4472 & 1.19 & $-$2.21 & 2.02 & 45 & $-$2.209 & 0.116 & 10 &  $-$2.226 & 0.106 \\ 
\hline
\end{tabular}
\label{t:atmpar08}
\end{table*}

\setcounter{table}{6}
\begin{table*}
\centering
\caption[]{Abundances of proton-capture elements in stars of Ter~8.
n is the number of lines used in the analysis. Upper limits (limNa,Al=0)
and detections (=1) for Na and Al are flagged.}  
\begin{tabular}{rrccrccrccrcccc}
\hline
       star  &n &  [O/Fe]&  rms  &  n& [Na/Fe]& rms  &	n& [Mg/Fe]& rms  &  n&[Al/Fe]& rms  &limNa&    limAl  \\ %
\hline       
  1209 &   &       &      & 2 &  +0.14 & 0.10 & 1 & 0.48 &      &   &       &	  & 1 &    \\ 
  1169 &   &       &      & 1 &  +0.05 &      & 1 & 0.37 &      &   &       &	  & 0 &    \\ 
  2086 &   &       &      & 2 &  +0.24 & 0.11 & 1 & 0.50 &      &   &       &	  & 1 &    \\ 
   526 &   &       &      & 1 &$-$0.10 &      &   &      &      &   &       &	  & 1 &    \\ 
  1728 &   &       &      & 2 &  +0.33 & 0.07 & 1 & 0.36 &      &   &       &	  & 0 &    \\ 
  2913 &   &       &      &   &        &      & 1 & 0.58 &      &   &       &	  &   &    \\ 
   514 &   &       &      &   &        &      & 1 & 0.43 &      &   &       &	  &   &    \\ 
   496 &   &       &      & 1 &  +0.10 &      & 1 & 0.36 &      &   &       &	  & 1 &    \\ 
   126 &   &       &      & 1 &  +0.10 &      & 1 & 0.47 &      &   &       &	  & 1 &    \\ 
  2124 &   &       &      & 1 &  +0.13 &      & 1 & 0.56 &      &   &       &	  & 0 &    \\ 
  1077 &   &       &      & 1 &$-$0.06 &      &   &      &      &   &       &	  & 1 &    \\ 
  2531 &   &       &      &   &        &      & 1 & 0.50 &      &   &       &	  &   &    \\ 
  2180 &   &       &      &   &        &      &   &      &      &   &       &	  &   &    \\ 
  2023 &   &       &      & 2 &  +0.88 & 0.01 &   &      &      &   &       &	  & 1 &    \\ 
  2357 & 1 & +0.44 &      & 3 &  +0.29 & 0.11 & 2 & 0.48 & 0.14 & 1 & +0.84 &	  & 1 & 0  \\ 
  1658 & 1 & +0.40 &      & 3 &  +0.19 & 0.04 & 2 & 0.51 & 0.02 & 1 & +0.77 &	  & 1 & 0  \\ 
   530 & 1 & +0.31 &      & 3 &  +0.25 & 0.07 & 2 & 0.60 & 0.26 & 1 & +0.85 &	  & 1 & 0  \\ 
  1188 & 1 & +0.42 &      & 3 &  +0.50 & 0.19 & 2 & 0.50 & 0.26 & 1 & +1.27 &	  & 1 & 0  \\ 
  2253 & 1 & +0.40 &      & 3 &  +0.15 & 0.14 & 2 & 0.33 & 0.21 & 1 & +0.94 &	  & 1 & 0  \\ 
   137 & 1 & +0.40 &      & 3 &  +0.14 & 0.07 & 2 & 0.40 & 0.16 & 1 & +1.09 &	  & 1 & 0  \\ 
\hline
\end{tabular}
\label{t:proton08}
\end{table*}

\clearpage
\setcounter{table}{7}
\begin{table*}
\centering
\caption[]{Abundances of $\alpha$-elements in stars of Ter~8. 
n is the number of lines used in the analysis.}
\begin{tabular}{rrccrccrccrcc}
\hline
   star      &  n&[Si/Fe]&  rms &    n &  [Ca/Fe]& rms &   n &[Ti/Fe]~{\sc i} &  rms &n &[Ti/Fe]~{\sc ii} & rms \\
\hline   
  1209 & 2 & +0.35 & 0.04 &  6 &  +0.21 & 0.22 &  4 &   +0.00 & 0.16 &   &         &      \\ 
  1169 &   &       &      &  5 &  +0.19 & 0.23 &  1 &   +0.14 &      &   &         &      \\ 
  2086 & 3 & +0.55 & 0.03 &  4 &  +0.21 & 0.04 &  2 &   +0.16 & 0.24 &   &         &      \\ 
   526 & 2 & +0.54 & 0.05 &  4 &  +0.33 & 0.24 &  2 &   +0.13 & 0.08 &   &         &      \\ 
  1728 & 2 & +0.42 & 0.06 &  5 &  +0.24 & 0.30 &  1 &   +0.06 &      &   &         &      \\ 
  2913 &   &       &      &  2 &$-$0.03 & 0.15 &    &         &      &   &         &      \\ 
   514 &   &       &      &  3 &  +0.19 & 0.26 &  2 &   +0.23 & 0.06 &   &         &      \\ 
   496 & 3 & +0.28 & 0.33 &  4 &  +0.21 & 0.13 &  1 & $-$0.09 &      &   &         &      \\ 
   126 &   &       &      &  4 &  +0.18 & 0.17 &  1 &   +0.12 &      &   &         &      \\ 
  2124 & 3 & +0.44 & 0.29 &  3 &  +0.26 & 0.20 &    &         &      &   &         &      \\ 
  1077 & 2 & +0.40 & 0.10 &  4 &  +0.35 & 0.28 &  2 &   +0.37 & 0.08 &   &         &      \\ 
  2531 &   &       &      &  4 &  +0.08 & 0.03 &  3 &   +0.12 &      &   &         &      \\ 
  2180 & 1 & +0.44 &      &  3 &  +0.13 & 0.13 &  1 &   +0.01 &      &   &         &      \\ 
  2023 &   &       &      &  3 &  +0.11 & 0.24 &    &         &      &   &         &      \\ 
  2357 & 1 & +0.25 &      & 18 &  +0.21 & 0.15 & 19 &   +0.07 & 0.11 & 6 &   +0.11 & 0.17 \\ 
  1658 & 1 & +0.24 &      & 15 &  +0.15 & 0.09 & 16 & $-$0.06 & 0.09 & 5 &   +0.18 & 0.14 \\ 
   530 & 2 & +0.24 & 0.11 & 18 &  +0.22 & 0.11 & 18 &   +0.06 & 0.07 & 7 &   +0.16 & 0.11 \\ 
  1188 & 1 & +0.32 &      & 13 &  +0.13 & 0.15 & 13 &   +0.06 & 0.15 & 6 &   +0.15 & 0.16 \\ 
  2253 & 1 & +0.08 &      & 19 &  +0.22 & 0.12 & 17 &   +0.05 & 0.15 & 6 & $-$0.01 & 0.12 \\ 
   137 & 1 & +0.35 &      & 15 &  +0.22 & 0.15 & 11 &   +0.12 & 0.09 & 5 &   +0.12 & 0.09 \\ 
\hline
\end{tabular}
\label{t:alpha08}
\end{table*}

\setcounter{table}{8}
\begin{table*}
\centering
\caption[]{Abundances of Fe-peak elements in stars of Ter~8. 
n is the number of lines used in the analysis.}
\scriptsize
\setlength{\tabcolsep}{1.3mm}
\begin{tabular}{rrccrccrccrccrccrccrcc}
\hline
      star    & n &[Sc/Fe]~{\sc ii}&rms&n& [V/Fe]  & rms  &  n &[Cr/Fe]~{\sc i}&rms&n& [Mn/Fe] & rms  &   n &[Ni/Fe] & rms   &  n  &[Cu/Fe]  & rms  &  n& [Zn/Fe] &rms\\
\hline         
  1209 & 7 & $-$0.07 & 0.11 & 5 & $-$0.11 & 0.22 & 1 & $-$0.15 &      &   &	    &	   &  5 & $-$0.05 & 0.10 &   &         &      &   &	    &	  \\ 
  1169 & 7 & $-$0.09 & 0.11 & 2 &   +0.00 & 0.13 & 1 & $-$0.03 &      &   &	    &	   &  2 & $-$0.05 & 0.10 &   &         &      &   &	    &	  \\ 
  2086 & 5 & $-$0.11 & 0.14 & 1 &   +0.04 &	 &   &         &      &   &	    &	   &  3 &   +0.03 & 0.31 &   &         &      &   &	    &	  \\ 
   526 & 5 &   +0.00 & 0.11 & 3 &   +0.07 & 0.11 &   &         &      &   &	    &	   &  2 &   +0.05 & 0.46 &   &         &      &   &	    &	  \\ 
  1728 & 6 & $-$0.08 & 0.35 & 2 &   +0.10 & 0.02 &   &         &      &   &	    &	   &  1 & $-$0.04 &	 &   &         &      &   &	    &	  \\ 
  2913 &   &	     &      &	&      99 &	 &   &         &      &   &	    &	   &  2 &   +0.21 & 0.02 &   &         &      &   &	    &	  \\ 
   514 & 5 &   +0.10 & 0.31 & 1 &   +0.13 &	 &   &         &      &   &	    &	   &  1 & $-$0.04 &	 &   &         &      &   &	    &	  \\ 
   496 & 7 &   +0.02 & 0.16 & 2 & $-$0.11 & 0.09 & 1 & $-$0.18 &      &   &	    &	   &  3 & $-$0.05 & 0.14 &   &         &      &   &	    &	  \\ 
   126 & 4 &   +0.06 & 0.19 &	&      99 &	 &   &         &      &   &	    &	   &  1 & $-$0.04 &	 &   &         &      &   &	    &	  \\ 
  2124 & 6 &   +0.05 & 0.29 &	&      99 &	 &   &         &      &   &	    &	   &  2 & $-$0.01 & 0.14 &   &         &      &   &	    &	  \\ 
  1077 & 4 & $-$0.12 & 0.20 & 2 & $-$0.12 & 0.29 &   &         &      &   &	    &	   &  1 & $-$0.01 &	 &   &         &      &   &	    &	  \\ 
  2531 & 2 &   +0.06 & 0.11 & 1 &   +0.06 &	 &   &         &      &   &	    &	   &  1 & $-$0.14 &	 &   &         &      &   &	    &	  \\ 
  2180 & 3 & $-$0.01 & 0.15 & 1 &   +0.03 &	 &   &         &      &   &	    &	   &  2 &   +0.04 & 0.28 &   &         &      &   &	    &	  \\ 
  2023 & 4 &   +0.13 & 0.14 &	&      99 &	 &   &         &      &   &	    &	   &  1 & $-$0.00 &	 &   &         &      &   &	    &	  \\ 
  2357 & 7 & $-$0.09 & 0.08 & 4 & $-$0.35 & 0.08 & 8 & $-$0.47 & 0.10 & 5 & $-$0.56 & 0.21 & 12 & $-$0.17 & 0.10 & 1 & $-$0.55 &      & 1 & $-$0.02 &	  \\ 
  1658 & 7 & $-$0.13 & 0.11 & 5 & $-$0.23 & 0.15 & 7 & $-$0.42 & 0.08 & 5 & $-$0.47 & 0.14 &  4 & $-$0.13 & 0.22 & 1 & $-$0.55 &      & 1 & $-$0.05 &	  \\ 
   530 & 8 & $-$0.11 & 0.09 & 4 & $-$0.30 & 0.09 & 7 & $-$0.36 & 0.12 & 4 & $-$0.47 & 0.17 &  8 & $-$0.17 & 0.13 & 1 & $-$0.75 &      & 1 & $-$0.15 &	  \\ 
  1188 & 5 & $-$0.11 & 0.05 & 3 & $-$0.39 & 0.20 & 7 & $-$0.47 & 0.11 & 3 & $-$0.60 & 0.24 &  5 & $-$0.21 & 0.17 & 1 & $-$0.59 &      & 1 &   +0.09 &	  \\ 
  2253 & 7 & $-$0.21 & 0.12 & 2 & $-$0.27 & 0.04 & 7 & $-$0.38 & 0.14 & 5 & $-$0.55 & 0.15 & 12 & $-$0.20 & 0.16 & 1 & $-$0.60 &      & 1 & $-$0.12 &	  \\ 
   137 & 5 & $-$0.07 & 0.04 & 1 & $-$0.26 &	 & 7 & $-$0.35 & 0.17 & 2 & $-$0.50 & 0.01 &  4 & $-$0.18 & 0.04 & 1 & $-$0.65 &      & 1 & $-$0.04 &	  \\ 
\hline
\end{tabular}
\label{t:fegroup08}
\end{table*}
\clearpage

\clearpage
\setcounter{table}{9}
\begin{table*}
\centering
\caption[]{Abundances of $n-$capture elements in stars of Ter~8 with 
UVES spectra; n is the number of lines used in the analysis.}
\scriptsize
\setlength{\tabcolsep}{1.3mm}
\begin{tabular}{rrccrccrccrccrccrccrcc}
\hline
  star    & n   &[Y/Fe]~{\sc ii}&rms&n& [Ba/Fe]~{\sc ii} & rms&n &[Nd/Fe]~{\sc ii} & rms   &  n  &[Eu/Fe]~{\sc ii}  & rms  \\
\hline         
2357 & 1 & $$0.10 &   & 3 & $-$0.13 & 0.04 & 2 & $$0.23 & 0.59    & 1 & $<$2.39 &	  \\ 
1658 & 1 & $$0.00 &   & 3 & $-$0.10 & 0.04 & 2 & $-$0.25 & 0.12   & 1 & $<$2.79 &	  \\ 
530  & 1 & $$0.00 &   & 3 & $-$0.19 & 0.03 & 4 & $-$0.55 & 0.12   & 1 & $<$2.63 &	  \\ 
1188 & 1 & $$0.05 &   & 3 & $-$0.05 & 0.06 & 3 & $-$0.57 & 0.17   & 1 & $<$2.70 &	  \\ 
2253 & 1 & $$0.20 &   & 3 & $-$0.26 & 0.05 & 4 & $-$0.52 & 0.12   & 1 & $<$2.59 &	  \\ 
137  & 1 & $$0.20 &   & 3 & $-$0.19 & 0.06 & 3 & $-$0.12 & 0.23   & 1 & $<$2.77 &         \\ 
1209 &   &        &   & 1 & $-$0.18 &      &   &         &        &   &         &         \\ 
1169 &	 &	  &   & 1 & $-$0.15 &	   &   &	 &	  &   & 	&	  \\ 
2086 &	 &	  &   & 1 & $-$0.22 &	   &   &	 &	  &   & 	&	  \\ 
 526 &	 &	  &   & 1 & $-$0.16 &	   &   &	 &	  &   & 	&	  \\ 
1728 &	 &	  &   & 1 & $-$0.61 &	   &   &	 &	  &   & 	&	  \\ 
2913 &	 &	  &   & 1 & $-$0.75 &	   &   &	 &	  &   & 	&	  \\ 
 514 &	 &	  &   & 1 & $-$0.30 &	   &   &	 &	  &   & 	&	  \\ 
 496 &	 &	  &   & 1 & $-$0.12 &	   &   &	 &	  &   & 	&	  \\ 
 126 &	 &	  &   & 1 &   +0.01 &	   &   &	 &	  &   & 	&	  \\ 
2124 &	 &	  &   & 1 & $-$0.12 &	   &   &	 &	  &   & 	&	  \\ 
1077 &	 &	  &   & 1 & $-$0.53 &	   &   &	 &	  &   & 	&	  \\ 
2531 &	 &	  &   & 1 & $-$0.93 &	   &   &	 &	  &   & 	&	  \\ 
2180 &	 &	  &   & 1 &   +0.30 &	   &   &	 &	  &   & 	&	  \\ 
2023 &	 &	  &   & 1 & $-$0.34 &	   &   &	 &	  &   & 	&	  \\ 
\hline
\end{tabular}
\label{t:neutron08}
\end{table*}

\end{document}